\documentclass[aos,preprint]{imsart}

\RequirePackage[OT1]{fontenc}
\RequirePackage{amsthm,amsmath}
\RequirePackage[numbers]{natbib}
\RequirePackage[colorlinks,citecolor=blue,urlcolor=blue]{hyperref}

\usepackage{latexsym, epsfig, amssymb, amsmath, amsthm, graphicx, mathrsfs}
\usepackage[usenames,dvipsnames]{color}

\startlocaldefs

\theoremstyle{plain}
\newtheorem{condition}{Condition}
\newtheorem{theorem}{Theorem}
\newtheorem{proposition}{Proposition}
\newtheorem{lemma}{Lemma}

\newcommand{\be}{\mbox{\bf e}}
\newcommand{\bff}{\mbox{\bf f}}

\newcommand{\bu}{\mbox{\bf u}}
\newcommand{\bv}{\mbox{\bf v}}
\newcommand{\bw}{\mbox{\bf w}}
\newcommand{\bx}{\mbox{\bf x}}
\newcommand{\by}{\mbox{\bf y}}

\newcommand{\bB}{\mbox{\bf B}}
\newcommand{\bC}{\mbox{\bf C}}
\newcommand{\bD}{\mbox{\bf D}}
\newcommand{\bE}{\mbox{\bf E}}
\newcommand{\bF}{\mbox{\bf F}}
\newcommand{\bG}{\mbox{\bf G}}

\newcommand{\bR}{\mbox{\bf R}}
\newcommand{\bS}{\mbox{\bf S}}

\newcommand{\bX}{\mbox{\bf X}}

\newcommand{\bY}{\mbox{\bf Y}}

\newcommand{\bone}{\mbox{\bf 1}}
\newcommand{\bzero}{\mbox{\bf 0}}
\newcommand{\bveps}{\mbox{\boldmath $\varepsilon$}}

\newcommand{\bbeta}{\mbox{\boldmath $\beta$}}

\newcommand{\bet}{\mbox{\boldmath $\eta$}}
\newcommand{\bxi}{\mbox{\boldmath $\xi$}}

\newcommand{\bmu}{\mbox{\boldmath $\mu$}}

\newcommand{\hbet}{\widehat \bet}

\newcommand{\hbbeta}{\widehat\bbeta}

\newcommand{\htheta}{\widehat\theta}
\newcommand{\hSig}{\widehat\Sig}
\newcommand{\hmu}{\widehat\bmu}

\newcommand{\hbx}{\widehat{\bx}}
\newcommand{\tbx}{\widetilde{\bx}}

\newcommand{\var}{\mathrm{var}}
\newcommand{\cov}{\mathrm{cov}}

\newcommand{\Sig}{\mathbf{\Sigma}}
\newcommand{\Ome}{\mathbf{\Omega}}
\newcommand{\ghOmeisee}{\widehat{\Ome}_{\mbox{\scriptsize ISEE}, g}}
\newcommand{\hOmeisee}{\widehat{\Ome}_{\mbox{\scriptsize ISEE}}}
\newcommand{\ihOmeisee}{\widehat{\Ome}_{\mbox{\scriptsize ISEE,ini}}}
\newcommand{\cihOmeisee}{\widehat{\Ome}_{\mbox{\scriptsize ISEE,cini}}}
\newcommand{\hEisee}{\widehat{E}_{\mbox{\scriptsize ISEE}}}
\newcommand{\hOme}{\widehat{\Ome}}
\newcommand{\veps}{\varepsilon}

\newcommand{\diag}{\mathrm{diag}}

\newcommand{\supp}{\mathrm{supp}}

\def\t{^T}

\def\boxit#1{\vbox{\hrule\hbox{\vrule\kern6pt\vbox{\kern6pt#1\kern6pt}\kern6pt\vrule}\hrule}}

\endlocaldefs

\begin{document}

\begin{frontmatter}
\title{Innovated Scalable Efficient Estimation in Ultra-Large Gaussian Graphical Models \thanksref{T1}}
\runtitle{ISEE}
\thankstext{T1}{
This work was supported by NSF CAREER Awards DMS-0955316 and DMS-1150318 and a grant from the Simons Foundation. The authors would like to thank the Co-Editor, Associate Editor, and referees for their valuable comments that have helped improve the paper significantly. Part of this work was completed while the authors visited the Departments of Statistics at University of California, Berkeley and Stanford University. The authors sincerely thank both departments for their hospitality.}

\begin{aug}
\author{\fnms{Yingying} \snm{Fan}\ead[label=e1]{fanyingy@marshall.usc.edu}}
\and
\author{\fnms{Jinchi} \snm{Lv}\ead[label=e2]{jinchilv@marshall.usc.edu}}

\runauthor{Y. Fan and J. Lv}

\affiliation{University of Southern California}

\address{Data Sciences and Operations Department\\
Marshall School of Business\\
University of Southern California\\
Los Angeles, CA 90089\\
USA\\
\printead{e1}\\
\phantom{E-mail:\ }\printead*{e2}}
\end{aug}

\begin{abstract}
Large-scale precision matrix estimation is of fundamental importance yet challenging in many contemporary applications for recovering Gaussian graphical models. In this paper, we suggest a new approach of innovated scalable efficient estimation (ISEE) for estimating large precision matrix. Motivated by the innovated transformation, we convert the original problem into that of large covariance matrix estimation. The suggested method combines the strengths of recent advances in high-dimensional sparse modeling and large covariance matrix estimation. Compared to existing approaches, our method is scalable and can deal with much larger precision matrices with simple tuning. Under mild regularity conditions, we establish that this procedure can recover the underlying graphical structure with significant probability and provide efficient estimation of link strengths. Both computational and theoretical advantages of the procedure are evidenced through simulation and real data examples.
\end{abstract}

\begin{keyword}[class=MSC]
\kwd[Primary ]{62H12}
\kwd{62F12}
\kwd[; secondary ]{62J05}
\end{keyword}

\begin{keyword}
\kwd{Gaussian graphical model}
\kwd{precision matrix}
\kwd{big data}
\kwd{scalability}
\kwd{efficiency}
\kwd{sparsity.}
\end{keyword}

\end{frontmatter}

\section{Introduction} \label{Sec1}
The surge of big data in an unprecedented scale has brought us an enormous amount of information about individuals in a spectrum of contemporary applications including social networks, online marketing, and modern healthcare. It is often of practical interest to uncover the underlying network formed by a large number of individuals that are sparsely related. Graphical models provide a flexible way to specify the conditional independence structure among a set of nodes. See, for example, \cite{Lauritzen96, WJ08} for detailed accounts and applications of such models. In Gaussian graphical models, the conditional independence structure is fully characterized by the zero entries in the precision (inverse covariance) matrix. For instance, the nonzero entries of a precision matrix estimated from genomic data detect interactions among genes or proteins of potential interest. The precision matrix also appears in many other applications such as classification and portfolio management.

The problem of identifying zeros in the precision matrix was termed as covariance selection in \cite{Dempster72}, which serves as a parsimonious way to simplify the model on the covariance structure. A stepwise 
estimation procedure was proposed therein based on the rule that the covariance matrix estimator is positive definite and matches the sample one on a set of entries, while its inverse has zeros in the remaining entries. In the Gaussian setting, it was shown that such a covariance model attains maximum entropy (simplicity) and the proposed covariance matrix estimator has the appealing property of being the restricted maximum likelihood estimate. 
Such a procedure works for the case when the number of variables $p$ is low but becomes computationally expensive as $p$ increases. 

Large precision matrix estimation has attracted much recent attention of many researchers. Broadly speaking, existing methods can be classified into two classes: the penalized likelihood or empirical risk methods, and the penalized regression or Dantzig selector type optimization methods. The former class includes, for example, \cite{YuanLin2007, FriedmanHastieTibshirani08, FanFengWu09, Ravikumaretal11, ZhangZou2014}. These methods share a common feature that the precision matrix is estimated by maximizing the penalized Gaussian likelihood or minimizing the penalized empirical risk. The latter class includes, for instance, \cite{MeinshausenBuhlmann06, Pengetal09, Yuan2010, CaiLiuLuo11, RSZZ13, CaiLiuZhou14}. Such methods convert the problem of precision matrix estimation into a nodewise or pairwise regression, or optimization problem and then apply the technique of high-dimensional regularization using the Lasso or Dantzig selector type methods. In particular, optimal rates of convergence for estimating sparse precision matrix have been established in \cite{CaiLiuZhou14}. The aforementioned methods are efficient in estimating precision matrix in moderate dimensions, but may become computationally inefficient when dealing with a huge number of nodes.

To address the important issue of scalability that is crucial to uncovering ultra-large Gaussian graphical models, in this paper we suggest a new method, called the innovated scalable efficient estimation (ISEE), for large precision matrix estimation. Our approach is motivated by the idea of the innovated transformation, which is a linear transformation of the $p$-variate random vector for the $p$ nodes using the precision matrix; see (\ref{006}) for formal definition. A simple observation is that the covariance matrix of the transformed $p$-variate random vector is exactly the precision matrix of the original $p$-variate random vector. Aided by such a transformation, we convert the original problem of large precision matrix estimation into that of large covariance matrix estimation. To estimate the innovated data matrix, the so-called oracle empirical matrix (see (\ref{008}) for formal definition), which is unavailable to practitioners, we exploit the scaled Lasso regression in \cite{SunZhang12} applied $p$ times based on a partition of all the $p$ nodes. After obtaining such a data matrix, we treat it as a ``sample" from the innovated $p$-variate random vector, and apply the approach of thresholding in \cite{BL08b} to construct a sparse precision matrix estimator.

The innovated transformation is related to the term ``innovation" used in the time series literature \cite{HallJin10} and has been utilized by other researchers in various contexts. For example, it was proposed and exploited in \cite{HallJin10} for detecting sparse signals when the noises are correlated. It was used in \cite{FJY13} for high-dimensional optimal classification with correlated features. See also \cite{jin2012comment} for a discussion of the innovated transformation in the multiple testing setting.

The suggested ISEE method combines the strengths of recent advances in both fields of high-dimensional sparse modeling and large covariance matrix estimation. The scaled Lasso is a convex regularization method that is tuning free and admits efficient implementation, while the thresholding method for large covariance matrix estimation is easy to implement and powered by appealing theoretical properties. As a consequence, there is only one tuning parameter for ISEE which is the threshold. To select such a threshold, we adapt the method of the cross-validation in \cite{BL08a, BL08b} for large covariance matrix estimation. Since we apply the cross-validation to the estimated oracle empirical matrix, not the original data matrix, there is no need to repeat the sparse regression step and thus the ISEE enjoys computational efficiency. As such, ISEE is scalable and can deal with much larger precision matrices with simple tuning, compared to existing approaches. 
In addition to the computational advantage, we have also shown that the suggested procedure can recover the underlying graphical structure with significant probability and provide efficient estimation of link strengths under mild regularity conditions. 

The rest of the paper is organized as follows. Section \ref{Sec2} introduces the suggested approach of ISEE for large Gaussian graphical models, and discusses its computation in large or ultra-large scale. We present the asymptotic efficiency of the new method in Section \ref{Sec4}. Section \ref{Sec3} details some examples of applications for our method. We provide several numerical examples in Section \ref{Sec5}. Section \ref{Sec6} discusses some extensions of the suggested method to a few settings. The proofs of some main results are relegated to the Appendix. Additional proofs of main results and technical details are provided in the Supplementary Material.

\section{Innovated scalable efficient estimation in ultra-large Gaussian graphical models}  \label{Sec2}

\subsection{Model setting} \label{Sec2.1}
Consider the Gaussian graphical model $G = (V, E)$ for a $p$-variate random vector
\begin{equation} \label{001}
\bx = (X_1, \cdots, X_p)\t \sim N(\bmu, \Sig),
\end{equation}
where $\bmu$ is a $p$-dimensional mean vector, $\Sig = (\sigma_{jk})$ is a $p \times p$ covariance matrix, and $G$ is an undirected graph associated with $\bx$ with $V = \{X_1, \cdots, X_p\}$ the set of vertices (or nodes) and $E = \{(j, k)\}$ the set of edges (or links) between the vertices. In this model, the lack of an edge $(j, k)$ between a pair of vertices $X_j$ and $X_k$ is characterized by the probabilistic property that these two components are independent conditional on the remaining $p - 2$ vertices. In other words, the existence of an edge amounts to conditional dependence between the two vertices given all other ones. Denote by $\Ome = (\omega_{jk})$ the precision matrix, that is, the inverse $\Sig^{-1}$ of the covariance matrix $\Sig$. It is well known in the Gaussian graphical model theory that there is an edge $(j, k)$ between a pair of vertices $X_j$ and $X_k$ if and only if the corresponding entry $\omega_{jk}$ of the precision matrix $\Ome$ is nonzero. See, for example, \cite{Lauritzen96} for a detailed account of graphical models. Such a characterization of the edge set shows that the problem of recovering the Gaussian graph $G$ is equivalent to recovering the support
\begin{equation} \label{002}
\supp(\Ome) = \{(j, k): \omega_{jk} \neq 0\} \ \text{ modulo symmetry},
\end{equation}
meaning the equivalence of links 
between nodes $j$ and $k$ in undirected graphs, of the precision matrix $\Ome$. In particular, the strength of each link $(j, k)$ is characterized by the magnitude of the corresponding entry $\omega_{jk}$.

Suppose $(\bx_i)_{i = 1}^n$ is an independent and identically distributed (i.i.d.) sample from the Gaussian graphical model (\ref{001}). Without loss of generality, assume that the mean vector $\bmu = \bzero$ throughout the paper. One natural and important question is how to efficiently recover the graphical structure and infer about the link strengths in large scale, that is, when the number of nodes $p$ is large compared to the sample size $n$. We will address this problem in the remaining part of the paper. 

\subsection{Innovated scalable efficient estimation} \label{Sec2.2}
Estimating the precision matrix $\Ome$ associated with the Gaussian graph $G$ is challenging even in moderate dimensionality $p$. Directly inverting the sample covariance matrix is infeasible since it is singular when $p>n$. To overcome this difficulty, various methods have been proposed. As discussed in the Introduction, a common limitation of these methods is that they are computationally intensive which can restrain their applications when estimating very large graphs.

To address these challenges, we propose a new procedure called the innovated scalable efficient estimation (ISEE) for effective and efficient large precision matrix estimation. The main idea of our approach is to convert the problem of estimating large precision matrix $\Ome$ to that of estimating large covariance matrix. Our method is motivated by the following linear transformation
\begin{equation} \label{006}
\widetilde{\bx} = \Ome \bx.
\end{equation}
Observe that the $p$-variate transformed random vector $\tbx$ in (\ref{006}) still has a Gaussian distribution with mean $\bzero$ and covariance matrix
\begin{equation} \label{007}
\cov(\widetilde{\bx}) = \Ome \cov(\bx) \Ome = \Ome \Sig \Ome = \Ome.
\end{equation}
Thus, if the transformed vector $\tbx$ were observable, then estimating the precision matrix $\Ome$ could be achieved by estimating the covariance matrix of the $p$-variate Gaussian random vector $\widetilde{\bx}$. 
Our new view of this problem naturally provides flexible alternative ways of Gaussian graph estimation powered by recent developments in large covariance matrix estimation. See, for example, \cite{BL08a, BL08b, CaiLiu2011, CaiYuan2012,FFL08,LamFan09,
Rothmanetal08}, among others.

The transformation (\ref{006}) with the precision matrix $\Ome$ is termed as innovation in the time series literature. We thus refer to (\ref{006}) as the innovated transformation and incorporate the word ``innovated" in the name of ISEE. As mentioned in the Introduction, such a transformation has also been used in other settings. The innovated transformation (\ref{006}) is, however, not directly applicable for large precision matrix estimation because the transformed vector $\tbx$ is unobservable. Estimating $\tbx$ by the two parts according to \eqref{006} is infeasible since it depends on the unknown precision matrix $\Ome$ which is our estimation target. We overcome this difficulty by breaking the long vector $\tbx$ into small subvectors and then estimating each one as a whole with the representation \eqref{006}, which we describe in details as follows.

We start with introducing some notation that will be used repeatedly in our presentation. For any subsets $A, B \subset \{1, \cdots, p\}$, denote by $\bx_A$ a subvector of $\bx$ formed by its components with indices in $A$, and $\Ome_{A,B} = (\omega_{jk})_{j \in A, k \in B}$ a submatrix of $\Ome$ with rows in $A$ and columns in $B$. We also use the shorthand notation $\Ome_{A}$ for $\Ome_{A,A}$ for convenience. Note that by the definition of $\tbx$, we can write the subvector $\tbx_A$ in the following form
\begin{equation}\label{def:xtilde}
\tbx_A 
= \Ome_{A,A}\bet_A,
\end{equation}
where $\bet_A=\bx_A + \Ome_{A,A}^{-1}\Ome_{A,A^c}\bx_{A^c}$ with $A^c$ the complement of set $A$.

The estimation of the two terms on the right hand side of \eqref{def:xtilde} is interrelated and can be achieved simultaneously and effectively through linear regression techniques. The essence of our proposal comes from
a simple yet useful fact in Gaussian graphical model theory. Recall that in the Gaussian graphical model (\ref{001}), it holds for any subset $A \subset \{1, \cdots, p\}$ that
\begin{equation} \label{003}
\bx_A|\bx_{A^c} \sim N(-\Ome_{A,A}^{-1} \Ome_{A, A^c} \bx_{A^c}, \Ome_{A,A}^{-1}).
\end{equation}
The conditional distribution (\ref{003}) suggests a multivariate linear regression model
\begin{equation} \label{004}
\bx_A = \bC_A\t\bx_{A^c} + \bet_A,
\end{equation}
where $\bC_A = -\Ome_{A^c, A} \Ome_{A,A}^{-1}$ is a matrix of regression coefficients, and $\bet_A$ is the vector of model errors which takes the form introduced in \eqref{def:xtilde} and 
 has a multivariate Gaussian distribution $ N(\bzero, \Ome_{A,A}^{-1})$.

The representation of the subvector $\bx_A$ in (\ref{004})  suggests that
regression techniques can be exploited to estimate the unknown subvector $\tbx_A$. To see this, let $\hbet_A$ be the residual vector obtained by using some regression technique to fit model \eqref{004}.
Then the unknown matrix $\Ome_{A,A}$ can be estimated as the inverse of the sample covariance matrix of the model residual vector $\hbet_A$. Denote by $\hOme_{A}$ the resulting estimator. Then we can estimate the subvector $\tbx_A$ in \eqref{def:xtilde} as $\hbx_A = \hOme_{A} \hbet_A$.

Let $(A_l)_{l = 1}^L$ be a partition of the index set $\{1, \cdots, p\}$, that is, $\bigcup_{l=1}^L A_l = \{1,\cdots, p\}$ and $ A_l \cap A_{m} = \emptyset$ for any $1\leq l \neq m \leq L$. Although the ideas of our approach are applicable to the case of general $|A_l|$, 
to simplify the presentation we focus our attention on the case of $|A_l| = 2$ when the number of nodes $p$ is even, and the case of $|A_l| = 2$ or $3$ when $p$ is an odd number.
Without loss of generality, throughout the paper we consider the specific partition  $A_l = \{2 l - 1, 2 l\}$ for $1 \leq l \leq L - 1$ and $A_L = \{2 L - 1, \cdots, p\}$ with $L = \lfloor p/2\rfloor$ the integer part of $p/2$. The ISEE repeats the above procedure for each $A_l$ with $1 \leq l \leq L$ to obtain estimated subvectors $\hbx_{A_l}$'s, and then stacks all these subvectors together to form an estimate $\hbx$ of the oracle innovated vector $\tbx = \Ome\bx$. By doing so, the problem of estimating the precision matrix based on the original vector $\bx$ reduces to that of estimating the covariance matrix based on the estimated transformed vector $\hbx$.

By its nature, the ISEE breaks large-scale precision matrix estimation into smaller-scale linear regression problems, each of which can be solved effectively and efficiently. Thanks to the scalability of ISEE, it has advantages over existing methods in estimating very large precision matrices. Detailed comparisons of ISEE with existing methods are given in Section \ref{Sec2.4}. 

\subsection{Estimation procedure by ISEE} \label{Sec2.3}
We now discuss in detail the implementation of the ISEE procedure.
To ease the presentation, we introduce some matrix notation. Denote by $\bX = (\bx_1, \cdots, \bx_n)\t$ the $n \times p$ data matrix. We refer to the innovated data matrix
\begin{equation} \label{008}
\widetilde{\bX} = (\widetilde{\bx}_1, \cdots, \widetilde{\bx}_n)\t = \bX \Ome
\end{equation}
as the oracle empirical matrix, which is unavailable to practitioners. Using matrix notation, the multivariate linear regression model (\ref{004}) can be written as
\begin{equation} \label{005}
\bX_A = \bX_{A^c} \bC_A + \bE_A,
\end{equation}
where  $\bX_A$ and $\bX_{A^c}$ are the submatrices of $\bX$ with columns in $A$ and its complement $A^c$, respectively, 
and $\bE_A$ is an $n \times |A|$ model error matrix with rows as i.i.d. copies of $\bet_A\t$. Then the corresponding submatrix $\widetilde{\bX}_A$ can be written as
\begin{align}\label{Xtilde_A}
\widetilde{\bX}_{A} & = (\bX \Ome)_{A} = \bX_{A} \Ome_{A,A} + \bX_{A^c} \Ome_{A^c, A} \\
\nonumber
& = (\bX_{A} + \bX_{A^c} \Ome_{A^c, A} \Ome_{A,A}^{-1}) \Ome_{A,A} = \bE_{A} \Ome_{A,A}.
\end{align}
The representation in (\ref{Xtilde_A}) provides the foundation for the estimation of the oracle empirical matrix $\widetilde{\bX}$.


Many existing methods can be used to fit the Gaussian linear regression model \eqref{004} and obtain the estimates for $\Ome_{A,A}$ and $\bet_A$. To avoid the issue of overfitting caused by high dimensionality, some kind of regularization, however, needs to be applied to control model complexity. There is a large body of literature on regularization methods; see, for example, \cite{Tibshirani96, FL01, FP04, Zou06, Zhang10, LF09, CandesTao07}, among many others.  See also \cite{FanLv13} for the connections and differences for a wide class of regularization methods in high dimensions, and \cite{Lv13} for characterizations of the impacts of high dimensionality in finite samples. For our implementation, we suggest to use the scaled Lasso method proposed in \cite{SunZhang12}. We opt to work with this method for two main reasons. First, scaled Lasso is a natural likelihood-based extension of the Lasso \cite{Tibshirani96} that is tuning free and admits efficient implementation; see (\ref{010}) for details about its tuning-free feature. The efficient implementation of scaled Lasso greatly reduces the computational cost of ISEE. Second, as seen from \eqref{def:xtilde}, we are interested in the prediction property (i.e., the estimation of $\bet_A$) instead of the variable selection property (i.e., the estimation of $\bC_A$) when fitting \eqref{004}. The sampling properties of scaled Lasso as revealed in \cite{SunZhang12} guarantee the accuracy in estimating $\Ome_{A,A}$ and $\bet_A$, and thus the scaled Lasso is sufficient for our purpose. We also remark that alternatively one can also exploit regularization methods for multivariate linear regression models instead of fitting one response at a time as in the scaled Lasso. 

For each node $j$ in the index set $A$, let us consider the univariate linear regression model for response $\bX_j$, which is the $j$th column of the data matrix $\bX$, given by the multivariate linear regression model (\ref{005})
\begin{equation} \label{009}
\bX_j = \bX_{A^c} \bbeta_j + \bE_j,
\end{equation}
where the $(p - |A|)$-dimensional vector $\bbeta_j$ is the column of the regression coefficient matrix $\bC_A$ corresponding to node $j$ and the $n$-dimensional error vector $\bE_j$ is the corresponding column of the error matrix $\bE_A$. In model (\ref{009}), node $j$ is regressed on all nodes in the complement set $A^c$. As mentioned before, in contrast to the conventional setting, our object of interest now is on the error vector $\bE_j$, instead of directly on the regression coefficient vector $\bbeta_j$. Thus we treat the regression coefficient vector $\bbeta_j$ as a nuisance parameter, and estimate it along with the error standard deviation using the penalized least squares with the scaled Lasso
\begin{equation} \label{010}
(\hbbeta_j, \htheta_j^{1/2}) = \arg\min_{{\small \bbeta} \in \mathbb{R}^{p - |A|}, \ \sigma \geq 0} \left\{\frac{\|\bX_j - \bX_{A^c} \bbeta\|_2^2}{2 n \sigma} + \frac{\sigma}{2} + \lambda \|\bbeta_*\|_1\right\},
\end{equation}
where $\bbeta_*$ is the Hadamard (componentwise) product of two $(p - |A|)$-dimensional vectors $\bbeta$ and $(n^{-1/2} \|\bX_k\|_2)_{k \in A^c}$ with $\bX_k$ the $k$th column of $\bX$, $\lambda \geq 0$ is a regularization parameter associated with the weighted $L_1$-penalty, and $\|\bv\|_q$ denotes the $L_q$-norm of a given vector $\bv$ for $q \geq 1$. Here the minimizer $\htheta_j^{1/2}$, which is over $\sigma$, provides an estimator of the error standard deviation $\theta_j^{1/2} = \var^{1/2}(\eta_j)$, where $\eta_j$ is a component of $\bet_A$ corresponding to node $j$. The tuning-free feature of the scaled Lasso is entailed by the fact that the theoretical choice of the regularization parameter $\lambda = C \{(2 \log p)/n\}^{1/2}$ with $C > 1$ some constant which can be made free of the noise level in the linear regression model; see \cite{SunZhang12} for more details. Hereafter we fix such a universal choice of $\lambda$ for scaled Lasso in (\ref{010}), and discuss an automatic empirical choice for $\lambda$, which is indeed tuning free, in Section \ref{Sec5.1}.  The use of the scale vector $(n^{-1/2} \|\bX_k\|_2)_{k \in A^c}$ amounts to rescaling each column of the design matrix $\bX_{A^c}$ to have $L_2$-norm $n^{1/2}$, matching that of the constant covariate $\bone$ for the intercept, which is standard in the studies for regularization methods.

Based on the regression step, for each node $j$ in the index set $A$ we define
\begin{equation} \label{011}
\widehat{\bE}_j = \bX_j - \bX_{A^c} \hbbeta_j \quad \text{ and } \quad \widehat{\bE}_{A} = (\widehat{\bE}_j)_{j \in A},
\end{equation}
where $\hbbeta_j$ is defined in (\ref{010}) and $\widehat{\bE}_{A}$ is an $n \times |A|$ matrix consisting of columns $\widehat{\bE}_j$ with nodes $j$ in the index set $A$. Clearly, the residual vector $\widehat{\bE}_j$ is a natural estimate of the error vector $\bE_j$ and thus $\widehat{\bE}_{A}$ is a natural estimate of the error matrix $\bE_{A}$. In view of (\ref{005}) and (\ref{004}), the $|A| \times |A|$ matrix $n^{-1} \widehat{\bE}_{A}\t \widehat{\bE}_{A}$ is a natural estimator of the error covariance matrix $\Ome_{A,A}^{-1}$. This observation motivates us to construct a natural estimator
\begin{equation} \label{012}
\widehat{\Ome}_{A} = (n^{-1} \widehat{\bE}_{A}\t \widehat{\bE}_{A})^{-1}
\end{equation}
for the principal submatrix $\Ome_{A,A}$ of the precision matrix $\Ome$ given by the index set $A$. 
These observations suggest a simple plug-in estimator $\widehat{\bE}_{A} \widehat{\Ome}_{A}$ for the unobservable submatrix $\widetilde{\bX}_{A}$ in \eqref{Xtilde_A}. 

When $A$ ranges over a partition $(A_l)_{l = 1}^L$ of the index set $\{1, \cdots, p\}$, the ISEE estimates the oracle empirical matrix $\widetilde{\bX}$ as the $n \times p$ matrix
\begin{equation} \label{014}
\widehat{\bX} = (\widehat{\bX}_{A_l})_{1 \leq l \leq L},
\end{equation}
where the submatrix of $\widehat{\bX}$ with columns in the index set $A_l$ is given by $\widehat{\bX}_{A_l} = \widehat{\bE}_{A_l} \widehat{\Ome}_{A_l}$ as constructed before. 
Then the ISEE proceeds as follows:
\begin{enumerate}
\item[a)] (\textit{Recovery of graph}) \ First calculate the initial ISEE estimator as the sample covariance matrix
    \begin{equation} \label{038}
    \ihOmeisee = n^{-1} \widehat{\bX}\t \widehat{\bX}.
    \end{equation}
   Then for a given threshold $\tau \geq 0$, define
    \begin{equation} \label{092}
    \ghOmeisee = T_\tau(\ihOmeisee),
    \end{equation}
    where $T_\tau(\bB) = (b_{jk} 1_{\{|b_{jk}| \geq \tau\}})$ denotes the matrix $\bB = (b_{jk})$ thresholded at $\tau$.
    Estimate the graphical structure $E$, the set of links, as $\hEisee = \supp(\ghOmeisee)$.

\smallskip

\item[b)] (\textit{Estimation of link strength}) \ For each link $(j, k)$ in the recovered graph $\hEisee$ with nodes $j$ and $k$ from different index sets $A_l$'s, update the corresponding entry of $\ghOmeisee$ as the off-diagonal entry of the $2 \times 2$ matrix $\widehat{\Ome}_{A_l}$ given in (\ref{012}) with $A_l$ replaced by $\{j, k\}$. This yields a refined sparse precision matrix estimator $\hOmeisee$ for the link strength.
\end{enumerate}
We refer to the former $\ghOmeisee$ as the ISEE estimator for the graph, and the latter $\hOmeisee$ as the ISEE estimator with refinement throughout the paper. In particular, it is easy to see that the principal submatrix of the initial ISEE estimator $\ihOmeisee$ given by each index set $A_l$ is simply the matrix $\widehat{\Ome}_{A_l}$ given in (\ref{012}).

The choice of the threshold $\tau$ in (\ref{092}) is important for practical implementation. We adapt the cross-validation method proposed in \cite{BL08a, BL08b} for large covariance matrix estimation. Specifically, we randomly split the sample of $n$ rows of the estimated oracle empirical matrix $\widehat{\bX}$ into two subsamples of sizes $n_1$ and $n_2$, and repeat this $N_1$ times. Denote by $\ihOmeisee^{1, \nu}$ and $\ihOmeisee^{2, \nu}$ the corresponding sample covariance matrices as defined in (\ref{038}) based on these two subsamples, respectively, for the $\nu$th split. The threshold $\tau$ can be chosen to minimize
\begin{equation} \label{094}
R(\tau) = N_1^{-1} \sum_{\nu = 1}^{N_1} \left\|T_\tau(\ihOmeisee^{1, \nu}) - \ihOmeisee^{2, \nu}\right\|^2,
\end{equation}
where $\|\cdot\|$ denotes the Frobenius norm of a given matrix.


\subsection{Comparisons with existing methods} \label{Sec2.4}
The ISEE is closely related to the methods of precision matrix estimation proposed in \cite{RSZZ13} and \cite{Yuan2010} in that all three methods are rooted in the regression formulation \eqref{004}. 
For each $1\leq j \neq k\leq p$, the ANT method in \cite{RSZZ13} estimates the $(j,k)$-entry of $\Ome$ using the off-diagonal entry of $\hOme_{A}$ defined in \eqref{012} with $A = \{j,k\}$. Thus ANT needs to conduct $O(p^2)$ scaled Lasso regressions and can become more computationally expensive for large $p$. Based on the observation that the $j$th column of $\Ome$ can be written as
$(\Ome_{A,A}, -\Ome_{A, A}\bC_A\t)\t$ with $A = \{j\}$ and $\bC_A$ defined in \eqref{004}, \cite{Yuan2010} proposed to exploit the Danzig selector \cite{CandesTao07} to estimate $\bC_A$ and used a similar method as in ISEE to estimate $\Ome_{A,A}$. So it is seen that both ISEE and ANT rely on the residual vector in the regression model \eqref{004}, while the method in \cite{Yuan2010} relies on both the residual vector and the regression coefficient vector $\bC_A$ whose estimation can suffer from the bias issue related to the Dantzig selector. In addition, the method in \cite{Yuan2010} requires to select a tuning parameter for each node and is thus more demanding in tuning.

The ISEE is also related to the neighborhood selection method in  \cite{MeinshausenBuhlmann06} and joint estimation method in \cite{Pengetal09} in the sense that all methods estimate the graph via Lasso-type regressions. The main difference between the methods in \cite{MeinshausenBuhlmann06} and \cite{Pengetal09} is that the former conducts $p$ nodewise Lasso regressions for graph recovery and needs tuning parameter selection for each node, while the latter exploits a single joint Lasso regression for precision matrix estimation with only one tuning parameter. Both methods in \cite{MeinshausenBuhlmann06} and \cite{Pengetal09} require the irrepresentable-type condition for consistent graph recovery which can become stringent in large precision matrix estimation. 
The Lasso regularization for precision matrix estimation has also been exploited in \cite{ZhangZou2014}, who proposed a Lasso penalized D-trace procedure in which the Lasso penalty is applied to a new quadratic loss with a positive-definiteness constraint. As a result, the obtained estimator enjoys the nice property of positive definiteness. The sparse recovery property was also established under the irrepresentable-type condition.

The CLIME \cite{CaiLiuLuo11} is another popularly used method for precision matrix estimation. It estimates the graphical structure node by node using a Dantzig selector type procedure. For each node, a tuning parameter needs to be selected. As pointed out in \cite{RSZZ13}, in order to ensure consistency in graph recovery CLIME needs an additional threshold that depends on the $L_1$-norm of the true precision matrix $\Ome$, which is unknown and can be large.

As mentioned in the Introduction, the penalized likelihood (e.g., \cite{YuanLin2007, FriedmanHastieTibshirani08, FanFengWu09, Ravikumaretal11}) is a group of widely used methods for precision matrix estimation. In general, these methods are not scalable due to the complexity of the likelihood function. Thus they can be computationally expensive when the scale of the problem becomes large.

In summary, compared to those existing methods, the ISEE enjoys easy tuning and is scalable. As shown later in Section \ref{Sec4}, it also has nice asymptotic properties under mild regularity conditions. We will also provide numerical comparisons of ISEE with some popularly used methods in Section \ref{Sec5}.

\subsection{Computation} \label{Sec2.6}
In the new era of big data, designing procedures with scalability is key to powering contemporary applications. The ISEE method is naturally scalable since the main computational cost comes from the construction of the estimate $\widehat{\bX}$ for the oracle empirical matrix $\widetilde{\bX}$. Such an $n \times p$ matrix is constructed by running $p$ penalized linear regression fittings. These univariate response problems and the use of different permutations of the set of nodes $\{1, \cdots, p\}$, which can help boost the power of detecting important links, are perfect for parallel and distributed computing. 
The nodes $j$ in the same index set $A$ can be allocated to a common processor. 
These computational advantages of ISEE make it ideal for cloud computing which becomes more prevalent nowadays, and thus appealing for uncovering ultra-large 
sparse graphs with big data.

\section{Asymptotic efficiency of innovated scalable efficient estimation} \label{Sec4}

\subsection{Technical conditions} \label{Sec4.1}

For the technical analysis, we focus on the class of $K$-sparse Gaussian graphs with spectrum constraint
\begin{equation} \label{015}
\mathcal{G}(M, K) = \left\{\Ome: \ \begin{array}{l}
\text{each row has at most $K$ nonzero off-diagonal} \\
\text{entries and } M^{-1} \leq \lambda_{\min}(\Ome) \leq \lambda_{\max}(\Ome) \leq M
\end{array}
\right\},
\end{equation}
where $K$ is some positive integer that can grow with dimensionality $p$, $M \geq 1$ is some constant, and $\lambda_{\min}(\cdot)$ and $\lambda_{\max}(\cdot)$ denote the smallest and largest eigenvalues of a given symmetric matrix, respectively. For each graph in class (\ref{015}), the number of links for each node is bounded by $K$ from above and the precision matrix $\Ome$ has bounded spectrum. A generalized concept of sparsity is considered in \cite{RSZZ13} to allow for the case when a portion of the links can be weak, that is, close to zero but not exactly zero. To simplify the technical presentation, we content ourselves with the class of $K$-sparse Gaussian graphs. For notational simplicity, all rates of convergence involving $\log p$ and probability bounds involving $p$ are understood implicitly with $p$ regarded as $\max(p, n)$. For each index set $S \subset \{1, \cdots, p\}$, denote by $\bu_S$ and $\bu_{S^c}$ the subvectors of $\bu \in \mathbb{R}^p$ with components in $S$ and its complement $S^c$, respectively.

\begin{condition} \label{Cond1}
The Gaussian graph (\ref{001}) belongs to class $\mathcal{G}(M, K)$ with $K \leq c_0 n/(\log p)$ for some sufficiently small constant $c_0 > 0$, the partition $(A_l)_{l = 1}^{L}$ satisfies $1 \leq \min_{l} |A_l| \leq \max_{l} |A_l| = O(1)$, and $\lambda = (1 + \varepsilon) \{2 \delta (\log p)/n\}^{1/2} = o(1)$ for any constants $\delta \geq 2$ and $\varepsilon > 0$.
\end{condition}

\begin{condition} \label{Cond2}
There exist some constants $0 \leq \alpha \leq 1/2$ and $\xi > 1$ such that the $L_\infty$-norm cone invertibility factor
\begin{equation} \label{047}
F_\infty = \inf\left\{\frac{\|\Sig \bu\|_\infty}{\|\bu\|_\infty}: \ \begin{array}{l}
\|\bu_{S^c}\|_1 \leq \xi \|\bu_S\|_1 \neq 0 \text{ for some } \\
S \subset \{1, \cdots, p\} \text{ with } |S| \leq O(K)
\end{array}
\right\}
\end{equation}
of the covariance matrix $\Sig = \Ome^{-1}$ satisfies $F_\infty^{-1} = O(K^\alpha)$.
\end{condition}

%

\begin{proposition} \label{Prop1}
For any $\Ome \in \mathcal{G}(M, K)$, it holds that $\inf\{\|\Sig \bu\|_\infty/\|\bu\|_\infty: \bu \neq \bzero\} \geq (K + 1)^{-1/2} M^{-1}$ with $\Sig = \Ome^{-1}$.
\end{proposition}

Condition \ref{Cond1} assumes the sparsity of the precision matrix and imposes an upper bound on the sparsity level $K$. The assumption of $\max_{l} |A_l| = O(1)$ is made to simplify the technical presentation and can be relaxed. Condition \ref{Cond2} puts a constraint on the cone invertibility factor $F_{\infty}$.   Proposition \ref{Prop1} above shows that the constant $\alpha$ in Condition \ref{Cond2} is indeed bounded from above by $1/2$.  See, for example, \cite{YZ10} and \cite{SunZhang12} for more discussions on the cone invertibility factors under various norms. We remark that only Conditions \ref{Cond1} and \ref{Cond2} are needed for the theoretical development of ISEE approach alone.


\subsection{Main results} \label{Sec4.2}

Our first theorem establishes the entrywise infinity norm estimation bound for the initial ISEE estimator.

\begin{theorem} \label{Thm1}
Assume that Conditions \ref{Cond1}--\ref{Cond2} hold and $K^{1 + \alpha} \lambda = o(1)$. Then with probability $1 - o\{p^{-(\delta - 2)}\}$ tending to one the initial ISEE estimator $\ihOmeisee$ in (\ref{038}) satisfies that
\begin{equation} \label{037}
\left\|\ihOmeisee - \Ome\right\|_\infty = O\left(K^\alpha \lambda\right),
\end{equation}
where $\|\cdot\|_\infty$ denotes the entrywise $L_\infty$-norm of a given matrix.
\end{theorem}

From the proof of Theorem \ref{Thm1} we see that the rate of convergence for the initial ISEE estimator is the maximum of two components $O\{\max(K\lambda^2, \lambda)\}$ and $O(K^\alpha \lambda)$, corresponding to the block-diagonal and off-block-diagonal entries of $\ihOmeisee$, respectively. Note that the block-diagonal entries are estimated directly from \eqref{012}, while most of the off-block-diagonal ones are estimated from the cross product terms $n^{-1}\widehat{\bX}_{A_k}\t\widehat{\bX}_{A_l}$ with $k\neq l$. The difference in the two estimation procedures results in the difference in two rates of convergence. Since it is assumed in Theorem \ref{Thm1} that $K^{1+\alpha}\lambda = o(1)$ with $\alpha \geq 0$, the rate of convergence $O(K^{\alpha}\lambda)$ dominates that of $O\{\max(K\lambda^2, \lambda)\}$, meaning that the block-diagonal entries are generally estimated more accurately than the off-block-diagonal ones. 

As introduced in Section \ref{Sec2.3}, we apply  thresholding to obtain the ISEE estimator for the graph $\ghOmeisee$ defined in \eqref{092}.
For each identified link $(j, k)$ in the recovered graph $\hEisee = \supp(\ghOmeisee)$, 
the ISEE estimator with refinement $\hOmeisee$ 
updates its corresponding entry as the off-diagonal entry of the $2 \times 2$ matrix $\widehat{\Ome}_{A_l}$ given in (\ref{012}) with $A_l=\{j, k\}$. The following theorem shows that both sparse precision matrix estimators $\ghOmeisee$ and $\hOmeisee$ enjoy nice asymptotic properties.

\begin{theorem} \label{Thm2}
Assume that the conditions of Theorem \ref{Thm1} hold and $\omega_0 = \min\{|\omega_{jk}|: (j, k) \in \supp(\Ome)\} \geq \omega^*_0 = C K^\alpha \lambda$ with $C > 0$ some sufficiently large constant. Then with probability $1 - o\{p^{-(\delta - 2)}\}$ tending to one, it holds simultaneously that
\begin{itemize}
\item[a)] \emph{(Graph recovery)} $\supp(\ghOmeisee) = \supp(\Ome)$ for any $\tau \in [c \omega^*_0, \omega_0 - c \omega^*_0]$ with $0 < c < 1/2$ some constant;

\item[b)] \emph{(Graph screening)} $\supp(\Ome) \subset \supp(\ghOmeisee)$ for threshold $\tau$ chosen by cross-validation \eqref{094} with $n_1/n_2$ bounded away from $0$ and $\infty$;

\item[c)] \emph{(Efficient estimation)}
    \begin{equation} \label{093}
\left\|\hOmeisee - \Ome\right\|_\infty = O\left(\lambda\right).
\end{equation}
\end{itemize}
\end{theorem}

The first part of results in Theorem \ref{Thm2} is more of theoretical interest since the quantities $\omega_0$ and $\omega^*_0$ are generally unknown in practice. The second part provides a theoretical backup for a fast practical approach to choosing threshold in large graph screening. Comparing \eqref{093} to \eqref{037}, it is seen that ISEE with refinement has an improved rate of convergence for precision matrix estimation when $\alpha > 0$. Such an improvement occurs because the off-block-diagonal entries are estimated more accurately in the refinement step. We remark that the bound in (\ref{093}) is obtained as $O\{\max(K\lambda^2, \lambda)\}$ which becomes $O(\lambda)$ since $K^{1+\alpha}\lambda = o(1)$ and $\alpha \geq 0$.

\subsection{A bias corrected initial ISEE estimator}
A comparison of the rates of convergence in Theorems \ref{Thm1} and \ref{Thm2} shows that the initial ISEE estimator is generally biased when $\alpha > 0$. As mentioned in the discussion after Theorem \ref{Thm1}, such a bias stems from the estimation of the off-block-diagonal entries of the precision matrix $\Ome$ using the cross product terms $n^{-1}\widehat{\bX}_{A_k}\t\widehat{\bX}_{A_l}$ with $k\neq l$. Motivated by the technical analysis of the initial ISEE estimator $\ihOmeisee$, we now define a bias corrected initial ISEE estimator $\cihOmeisee$ as
\begin{equation} \label{bcisee01}
\big(\cihOmeisee\big)_{A_l, A_l} = \big(\ihOmeisee\big)_{A_l, A_l}
\end{equation}
and
\begin{equation} \label{bcisee02}
\big(\cihOmeisee\big)_{A_l, A_m} = -\big[\big(\ihOmeisee\big)_{A_l, A_m} + \widehat{\bC}_{A_l}^ {A_m} \widehat{\Ome}_{A_l} + \widehat{\bC}_{A_m}^ {A_l} \widehat{\Ome}_{A_m}\big]
\end{equation}
for each $1 \leq l \neq m \leq L$, where $\widehat{\bC}_{A_l} = (\hbbeta_{j,l})_{j \in A_l}$ represents a $(p - |A_l|) \times |A_l|$ matrix of estimated regression coefficients with $\hbbeta_{j,l}$ as defined in (\ref{010}), $\widehat{\bC}_{A_l}^ {A_m}$ denotes a submatrix of $\widehat{\bC}_{A_l}$ consisting of rows with indices in $A_m$, and $\widehat{\Ome}_{A_l}$ is given in (\ref{012}). The following theorem shows that such a bias corrected precision matrix estimator indeed admits improved rate of convergence.

\begin{theorem} \label{Thm5}
Under the conditions of Theorem \ref{Thm1}, the bias corrected initial ISEE estimator $\cihOmeisee$ in (\ref{bcisee01})--(\ref{bcisee02}) satisfies with probability $1 - o\{p^{-(\delta - 2)}\}$ tending to one that
\begin{equation} \label{112}
\left\|\cihOmeisee - \Ome\right\|_\infty = O\left(\lambda\right),
\end{equation}
and graph recovery consistency in part a of Theorem \ref{Thm2}, with $\ghOmeisee = T_\tau(\cihOmeisee)$ and $\omega^*_0 = C \lambda$ for some sufficiently large constant $C > 0$.
\end{theorem}

In light of Theorems \ref{Thm1}--\ref{Thm5}, 
we see that both the ISEE estimator with refinement $\hOmeisee$ and the bias corrected initial ISEE estimator $\cihOmeisee$ enjoy the same rate of convergence which is generally faster than that for the initial ISEE estimator $\ihOmeisee$ when $\alpha > 0$. 
We remark that our bias corrected initial ISEE estimator for the case of $|A_l| = 1$ for each $1 \leq l \leq L$ shares similar flavor to the bias corrected test statistics introduced in \cite{liu2013graph} for false discovery rate control in Gaussian graphical model estimation. In addition to the consistency result in Theorem \ref{Thm5}, we can further show that the estimates for zero entries of the precision matrix $\Ome$ can enjoy the asymptotic normality as in \cite{liu2013graph}. Due to space limitation, we do not pursue that direction in our current paper.

\section{Applications of innovated scalable efficient estimation} \label{Sec3}

As a byproduct, the ISEE procedure also provides a fast approach to estimating the innovated transformation \eqref{006}, which is key to methods such as the ideas of multiple testing using the innovated higher criticism in \cite{HallJin10}, the optimal classification in sparse Gaussian graphic models in \cite{FJY13}, and the interaction screening in high-dimensional quadratic discriminant analysis in \cite{FKLZ2015}. With the aid of ISEE, these methods can be more effectively and efficiently applied for the analysis of big data. We next discuss some additional applications of ISEE.

\subsection{Dimension reduction} \label{Sec3.3}

Dimension reduction facilitates greatly large-scale data analysis by effectively reducing the 
intrinsic dimensions of the feature space. 
Among all dimension reduction approaches, the sliced inverse regression (SIR) \cite{Li91} has been widely used. 
The SIR is based on the model
\begin{equation} \label{106}
Y = m(\bbeta_1\t \bx, \cdots, \bbeta_{K_0}\t \bx, \veps),
\end{equation}
where $Y$ is the response variable, $\bx$ is a $p$-dimensional covariate vector, $\bbeta_1, \cdots, \bbeta_{K_0}$ are unknown projection vectors with $1 \leq K_0 < p$ an unknown integer, $m: \mathbb{R}^{K_0 + 1} \rightarrow \mathbb{R}$ is an unknown function, and $\veps$ is the noise random variable with $E(\veps|\bx) = 0$. SIR aims at estimating the effective dimension reduction (EDR) space spanned by the EDR directions $\bbeta_k$'s \cite{Li91, HallLi93}. The SIR algorithm begins with standardizing the covariate vectors by centering and rescaling using the square-root precision matrix $\Ome^{1/2}$ of covariates, and produces an estimate of the EDR directions by multiplying the constructed eigenvectors by the same matrix.

The square-root precision matrix of covariates used in the SIR algorithm can be difficult and computationally expensive to estimate when $p$ is much larger than $n$. This problem can be resolved using the innovated transformation $\widetilde{\bx} = \Ome \bx$. 
Observe that by Theorem 3.1 in \cite{Li91}, the covariance matrix of $E(\widetilde{\bx}|Y)$ is degenerate in any direction orthogonal to the linear subspace spanned by the $K_0$ vectors $\cov(\widetilde{\bx}) \Sig \bbeta_k = \Ome \Sig \bbeta_k = \bbeta_k$, by noting that $\Sig \bbeta_k$ are the EDR directions for the transformed data $\tbx_1, \cdots, \tbx_n$. This suggests that the original EDR directions $\bbeta_k$'s can be obtained by calculating the eigenvectors of $\cov\{E(\tbx|Y)\}$. Since the oracle empirical matrix $\widetilde{\bX} = (\widetilde{\bx}_1, \cdots, \widetilde{\bx}_n)\t$ can be estimated effectively and efficiently using the idea of ISEE, $\cov\{E(\tbx|Y)\}$ can be estimated easily and thus this alternative approach greatly reduces the computational cost of large-scale dimension reduction using SIR.


\subsection{Portfolio management} \label{Sec3.4}

The precision matrix also plays an pivotal role in optimal portfolio allocation. Let $\by_i = \bmu + \bx_i$ be the return vector of $p$ assets at time $i$, where $\bmu\in \mathbb{R}^p$ is the vector of mean returns of the $p$ assets. Then $\Sig = \Ome^{-1}$ is the covariance matrix of these $p$ assets. Markowitz's mean-variance optimal portfolio \cite{Markowitz1952} is defined as the solution to  the following minimization problem:
\begin{equation}
\min_{\bxi \in \mathbb{R}^p} \bxi\t\Ome^{-1}\bxi \qquad \text{ subject to } \bxi\t\bone = 1 \text{ and } \bxi\t\bmu = \gamma,
\end{equation}
where $\bone$ is a $p$-vector of ones and $\gamma >0$ is the  targeted return imposed on the portfolio $\bxi$. It is well known that Markowitz's optimal portfolio admits an explicit solution
\[
\bxi_{\text{opt}} = \frac{d_1 - \gamma d_2}{d_3d_1 - d_2^2}\Ome\bone + \frac{\gamma d_3 - d_2}{d_3d_1 - d_2^2}\Ome\bmu,
\]
where $d_1 = \bmu\t\Ome\bmu$, $d_2 = \bone\t\Ome\bmu$, and $d_3 = \bone\t\Ome\bone$. With the ISEE estimate of the precision matrix $\Ome$, the optimal portfolio from a large number of assets can be easily constructed.
%

\subsection{Multiple testing, feature screening, and simultaneous confidence intervals} \label{Sec3.2}
Testing the significance of coefficients in a regression model is of particular importance in high dimensions, where feature selection is of interest in many applications. For simplicity, consider the linear regression model
\begin{equation} \label{102}
\by = \bX \bbeta + \bveps,
\end{equation}
where $\by$ is an $n$-vector of response, 
$\bbeta = (\beta_1, \cdots, \beta_p)\t$ is a $p$-vector of regression coefficients, and $\bveps$ is an $n$-vector of i.i.d. random error with variance $\sigma^2$. There is a large literature on multiple testing with the false discovery rate (FDR) control \cite{BH95}. It has been a convention to consider $p$ marginal regression models and test each of the $p$ marginal regression coefficients is equal to zero simultaneously. For example, \cite{FHG12} proposed the PFA method for high-dimensional multiple testing where the test statistics can have an arbitrary dependence structure. In contrast to testing the marginal effects of covariates, it is also interesting to test their joint effects
\begin{equation} \label{103}
H_{0j}: \ \beta_j = 0 \quad \text{versus} \quad H_{1j}: \ \beta_j \neq 0, \quad j = 1, \cdots, p.
\end{equation}

With the aid of the innovated transformation (\ref{006}), the multiple testing problem (\ref{103}) can be reduced to the scenario of marginal regression models linking the response and each of the $p$ innovated covariates. To see this, note that
\begin{equation} \label{104}
n^{-1}\widetilde\bX\t \by = n^{-1}\Ome \bX\t \by = \bbeta + \widetilde{\bveps},
\end{equation}
where $\widetilde{\bX} = \bX \Ome$ is the oracle empirical matrix and $\widetilde{\bveps} = - (I_p - n^{-1}\Ome \bX\t \bX) \bbeta + n^{-1}\Ome \bX\t \bveps$. Observe that $n^{-1}\bX\t\bX$ is the sample estimate of the covariance matrix $\Sig = \Ome^{-1}$. Thus intuitively $I_p - n^{-1}\Ome \bX\t \bX$ can be of a small order and thus the first term of $\widetilde{\bveps}$ can also be of a small order. Similarly the second term $n^{-1}\Ome \bX\t \bveps$ has mean $\bzero$ and conditional covariance matrix  $\cov(n^{-1}\Ome \bX\t \bveps|\bX) = n^{-2}\sigma^2\Ome\bX\bX\t\Ome$ which can be in the order of $n^{-1}\sigma^2 \Ome$.  Therefore, $\widetilde{\bveps}$ can be treated similarly as a random error vector.
The empirical version of (\ref{104}) can be obtained by substituting $\widetilde\bX$ with the estimate $\widehat\bX$ in \eqref{014}. The use of the innovated transformation has also been discussed in \cite{jin2012comment} to improve the performance of multiple testing using the correlation structure.

Feature screening with independence learning has been popularly used in both regression and classification problems. See, for example, \cite{FF08, HTX09, FL08}, among many others. Intuitively, it is natural and appealing to exploit the joint information among the covariates. The innovated features given by the ISEE estimator pool such joint information and provide new features that can be used for ranking the importance of original features, as elucidated in (\ref{104}). With the representation (\ref{104}), one can also construct simultaneous confidence intervals for the $p$ regression coefficients $\beta_j$'s using asymptotic distributions or the bootstrap \cite{Efron79}. 

\section{Numerical Studies} \label{Sec5}

\subsection{Implementation of ISEE} \label{Sec5.1}
When implementing ISEE, we choose the regularization parameter $\lambda$ in scaled Lasso following the suggestion of  \cite{RSZZ13}; that is, we fix $\lambda$ to be $B/(n-1+B^2)^{1/2}$, where $B = tq(1-n^{1/2}/(2p\log p), n-1)$ with $tq(\alpha, m)$ the $\alpha$th quantile of a $t$-distribution with $m$ degrees of freedom. The threshold $\tau$ is chosen adaptively using the random split method described in Section \ref{Sec2.3}. In both our simulation study and real data analysis, we use 90\% of the sample to calculate $\ihOmeisee^{1, \nu}$ and remaining 10\% to calculate $\ihOmeisee^{2, \nu}$, and then select $\tau$ from a grid of 20 values by minimizing the criterion \eqref{094} with the number of random splits set to $N_1 = 5$. Although the change in the computational cost of ISEE is negligible for a larger value of $N_1$ as discussed before, our choice of $N_1$ works well in empirical studies.

In our numerical studies, we observe that the ISEE estimator for the graph $\ghOmeisee$ calculated using the above way of tuning tends to have the number of false positives very close to zero, while the number of false negatives, which can also be close to zero, tends to be slightly larger than the number of false positives. Since the first step of ISEE focuses on recovering the underlying graph, the sure screening property, that is, zero false negative and low false positives with significant probability, is desirable for this step. To reduce the number of false negatives, we borrow idea from the Bonferroni method. Specifically, we first randomly permute the columns of the $n \times p$ data matrix $\bX$, then apply ISEE to the permuted data matrix to construct a sparse precision matrix estimator, and finally permute this sparse estimator back to obtain an estimate $\ghOmeisee^\pi$ of the original precision matrix, where $\pi$ denotes the corresponding permutation of $\{1, \cdots, p\}$ used in the estimate. We repeat this procedure $N_2$ times and construct the final estimate for the set of links of the graph as the union of $\supp(\ghOmeisee^\pi)$ over all $N_2$ permutations. For each identified link $(j, k)$, we average all the nonzero estimates of $\omega_{jk}$ over the $N_2$ repetitions to construct its final estimate. Although this permutation method adds to the computational cost of ISEE, it reduces the number of false negatives in all our settings. Moreover, thanks to the efficiency of ISEE for each fixed permutation the computational cost of our procedure is still much lower than those of other comparison methods even after we include this additional step.

We finally remark that in the simulation study, for a fair comparison with other methods ISEE is implemented without the refinement step. Thus the ISEE estimator in our simulation examples refers to $\ghOmeisee$ with the aforementioned way of tuning. 

\subsection{Simulation examples} \label{Sec5.2}

\subsubsection{Simulation example 1} \label{Sec5.2.1}
We start with a simulation example designed to compare the computational cost and accuracy of ISEE with some popularly used methods. We generate the precision matrix $\Ome$ in two steps. First, we produce a band matrix $\Ome_0$ with diagonal entries being one, $\Ome_0(i,i+1) = \Ome_0(i+1,i) = 0.5$ for $i = 1, \cdots, p-1$, and all other entries being zero. Second, we randomly permute the rows and columns of $\Ome_0$ to obtain the precision matrix $\Ome$. Thus in general, the final precision matrix $\Ome$ no longer has the band structure. We then sample the rows of the $n \times p$ data matrix $\bX$ as i.i.d. copies from the multivariate Gaussian distribution $N(\bzero, \Ome^{-1})$. Throughout the simulation, we fix the sample size $n= 200$ and consider a range of dimensionality $p$.

The methods for comparison include the Glasso \cite{FriedmanHastieTibshirani08}, CLIME \cite{CaiLiuLuo11}, and ANT \cite{RSZZ13}. To implement Glasso and CLIME, we use the R packages \verb|glasso| and \verb|scio|, respectively. Both Glasso and CLIME have one tuning parameter, which is selected using fivefold cross-validation from a grid of 10 values. 
The ANT is implemented using the R package \verb|ConditionalGGM| with the tuning parameters set to the default values, and is thus tuning free. Our ISEE approach is implemented in the way described in Section \ref{Sec5.1}.

\begin{figure}[ht!] \centering
\begin{center}%
\begin{tabular}
[c]{c}%
{\hspace{-0.55in} \includegraphics[trim = 0mm 5mm 5mm 5mm, clip, width=12cm]%
{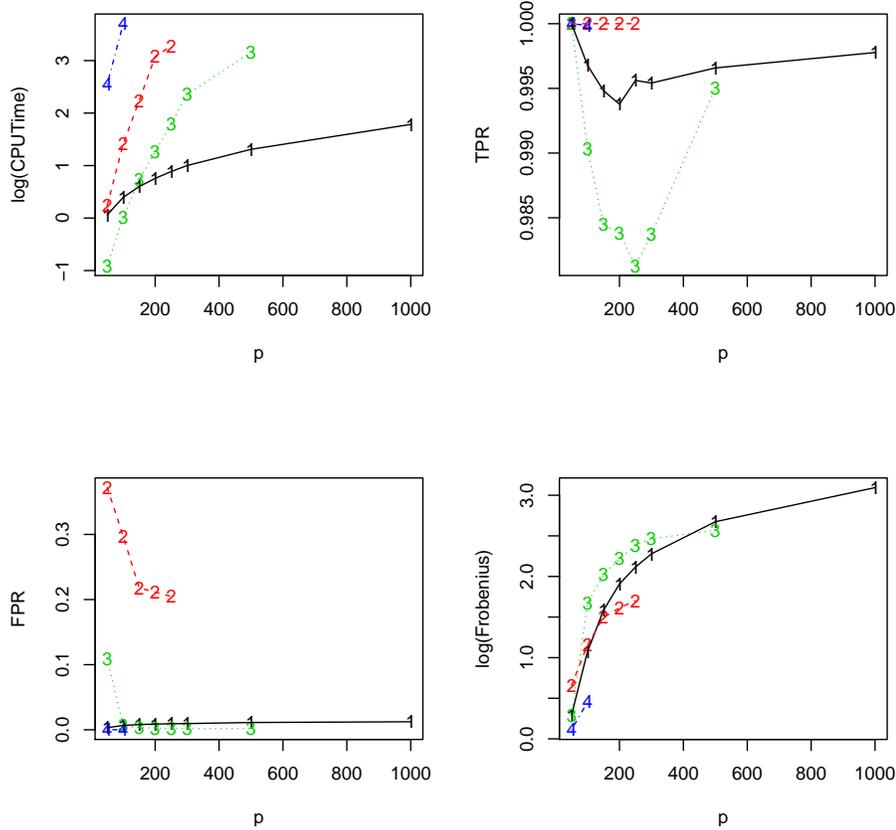}%
}%
\end{tabular}%
\caption{Comparisons of ISEE (marked as ``1"), Glasso (marked as ``2"), CLIME (marked as ``3"), and ANT (marked as ``4") in simulation example 1. Top left:  the common logarithm of CPU time as a function of dimensionality $p$; Top right: TPR as a function of dimensionality $p$; Bottom left: FPR as a function of dimensionality $p$; Bottom right: the common logarithm of estimation error under the Frobenius norm as a function of dimensionality $p$.}
\label{CPUTime}%
\end{center}%
\end{figure}%

We generate 50 data sets. For each data set, we run the four comparison methods on a PC with 8GB ram and Intel(R) Core(TM) i5-2500 CPU (3.30GHz). For each method in each repetition, the CPU time (in seconds) of obtaining a sparse precision matrix estimator is recorded. Three additional performance measures: the true positive rate (TPR), false positive rate (FPR), and estimation error under the Frobenius norm are also calculated. Here, the TPR and FPR are defined as
\begin{align*}
&\text{TPR} = \frac{\# \text{ of correctly identified edges}}{\text{ \# of identified edges in total}}, \\
& \text{FPR} = \frac{\# \text{ of falsely identified edges}}{\text{ \# of identified nonedges in total}},
\end{align*}
respectively.


The comparison results are summarized in Figure \ref{CPUTime}, with the $x$-axis indicating dimensionality $p$ and $y$-axis showing the mean values of different performance measures over 50 repetitions.  To make it easier to view, CPU time (top left) and estimation error under the Frobenius norm (bottom right) are both plotted under the common logarithmic scale. Due to high computational cost, the largest values of $p$ in our simulation for Glasso, CLIME, and ANT are 250, 500, and 100, respectively. It is seen that ISEE is computationally much more efficient than all other methods. CLIME is the second best in terms of CPU time. When $p=500$, ISEE is about 70 times faster than CLIME on average. Moreover, the accuracy of ISEE in support recovery and estimation is also among the best.

\subsubsection{Simulation example 2} \label{Sec5.2.2}
We now test the performance of ISEE in larger scales. We generate the precision matrix in two steps. First, we create a block-diagonal matrix $\Ome_0$ whose diagonal blocks are matrices of size 20. The diagonal entries of $\Ome_0$ are all equal to one. For each of the block matrix, the off-diagonal entries take value 0.5 with probability 0.3 and value 0 with probability 0.7. Since the matrix generated in this way may not be symmetric or positive definite, we first symmetrize it by forcing the lower triangular matrix to equal the upper triangular matrix, and then add a diagonal matrix $c I_{20}$ for some quantity $c$ to make the smallest eigenvalue of each block matrix equal to 0.1, where $I_{20}$ denotes an identity matrix of size 20. It is worth mentioning that in our example, the diagonal block matrices are generated independently of each other and are thus generally different. Second, we randomly permute the rows and columns of $\Ome_0$ to construct the final precision matrix $\Ome$. Thus, our true precision matrix $\Ome$ no longer has the block-diagonal structure. We consider two settings of dimensionality $p= 1000$ and $2000$, with the same sample size $n$ as in simulation example 1.

The same performance measures as in simulation example 1 are employed to evaluate the performance of ISEE. The means and standard errors over 100 repetitions are presented in Table \ref{tab1}, with Frob representing estimation error under the Frobenius norm. It is seen that even for these very large precision matrices, ISEE is still computationally efficient and performs well in graph recovery.

\begin{table}[t]
\caption{Performance of ISEE in simulation example 2.
\label{tab1}}
\begin{tabular}{clcccc}
\hline
 $p$     &       & Frob  & TPR   & FPR   & CPU Time \\
      \hline
$1000$ & Mean  & 3206.09 & 0.96799 & 0.05005 & 649.588 \\
      & SE    & 2.24128 & 0.00030 & 0.00006 & 0.70461 \\
\hline
$2000$ & Mean  & 7272.65 & 0.95867 & 0.03344 & 2287.34 \\
      & SE    & 3.42452 & 0.00023 & 0.00003 & 1.47541 \\
\hline
\end{tabular}%
\end{table}

%
%
%
%

%
%
%
%

\subsection{Real data analysis} \label{Sec5.3}
We finally evaluate the performance of ISEE on a breast cancer data set analyzed in \cite{Hess2006}. This data set consists of 22,283 gene expression levels of 130 breast cancer patients, among whom 33 patients had pathological complete response (pCR) and the remaining did not achieve pCR. Here, pCR is defined as no evidence of viable, invasive tumor cells left in the surgical specimen, and thus has been regarded as a strong indicator of survival.

This breast cancer data set has been used in \cite{CaiLiuLuo11} and \cite{FanFengWu09} to evaluate the accuracy of precision matrix estimation methods. We follow the steps therein to demonstrate the performance of ISEE. For completeness, we briefly list the data analysis procedure here. We first randomly split the data into training and test sets of sizes 109 and 21, respectively. Since the two classes have unbalanced sample size, a stratified sampling is used with 16 subjects randomly selected from pCR class and 5 subjects randomly selected from the other class to form the test set; the remaining subjects are used as the training set. Based on the training set, we conduct a two sample $t$-test and select the most significant $p=400$ genes with the smallest $p$-values. We remark that both \cite{CaiLiuLuo11} and \cite{FanFengWu09} kept only the most significant 110 genes. Thanks to the scalability of ISEE, we are able to deal with much larger precision matrix. We next conduct a gene-wise standardization by dividing the data matrix by the corresponding standard deviations. Then we estimate the $p\times p$ precision matrix $\Ome$ using the ISEE approach based on the training set, and construct the linear discriminant analysis (LDA) rule on the test set. The LDA assumes that both classes have Gaussian distributions $N(\bmu_k, \Ome^{-1})$ with different mean vectors $\bmu_1, \bmu_2$ and a common covariance matrix $\Sig = \Ome^{-1}$. With the ISEE estimator $\hOmeisee$ of the precision matrix, the discriminant function takes the form
\begin{equation}\label{LDA}
L(\bx) = (\bx\t - \bar{\bmu}) \hOmeisee \hmu + \log(n_1/n_2),
\end{equation}
where $\bar{\bmu} = (\hmu_1 + \hmu_2)/2$ and $\hmu = \hmu_1 - \hmu_2$ with $\hmu_k$, $k=1,2$, the sample mean vectors, and $n_1$ and $n_2$ are the training sample sizes from classes 1 (pCR) and 2, respectively. For a new observation vector $\bx$, LDA assigns it to class 1 if $L(\bx) > 0$ and to class 2 otherwise. Such a procedure is repeated 100 times.

As pointed out and done in \cite{CaiLiuLuo11}, an additional refit step may improve the accuracy of precision matrix estimation. We follow their suggestion and exploit a refitted ISEE estimator in calculating \eqref{LDA}. There are different ways to refit the ISEE estimator. One option is the ISEE estimator with refinement described in Section \ref{Sec2.3}. This approach of refitting can potentially suffer from growing computational cost for less sparse precision matrices. In our application, we adopt the refitting procedure described in \cite{FJY13}.  The main idea is to refit the ISEE estimator for the graph column by column after obtaining the support. Taking the first column as an example, ideally we would like to have
\begin{equation}\label{1st-col}
\hSig \hOme(,1) = \be_1,
\end{equation}
where $\hSig$ is the sample covariance matrix, $\hOme(,1)$ denotes the first column of a precision matrix estimator $\hOme$, and $\be_1$ is a $p$-vector with one in the first component and zero otherwise. Denote by $\mathcal{S} = \supp\{\hOme(,1)\}$ the recovered support of the first column. Then it follows from \eqref{1st-col} that
\begin{equation}
 \hSig_{\mathcal{S},\mathcal{S}}\hOme_{\mathcal{S},\mathcal{S}}(,1) = \be_{1,\mathcal{S}}.
 \end{equation}
Thus we can refit on the support $\mathcal{S}$ by inverting the principal submatrix $\hSig_{\mathcal{S},\mathcal{S}}$ and taking out the first column, that is, $(\hSig_{\mathcal{S},\mathcal{S}})^{-1}\be_{1,\mathcal{S}}$. Recall that in this paper, we consider the class of sparse precision matrices $\mathcal{G}(M, K)$ with $K = O\{n/(\log p)\}$. As guaranteed by Theorem \ref{Thm2}, ISEE enjoys nice graph recovery property and thus the size of the support $\mathcal{S}$ can be much smaller than the sample size $n$ with significant probability. So generally the inverse of the matrix $\hSig_{\mathcal{S},\mathcal{S}}$ can be  obtained efficiently. Nevertheless, to enhance stability in real applications we suggest the use of the generalized inverse of matrix $\hSig_{\mathcal{S},\mathcal{S}}$ if $|\mathcal{S}|$ is close to or exceeds $n$. 

To evaluate the performance of classification rule (\ref{LDA}), we consider three measures: the specificity, sensitivity, and Matthews correlation coefficient (MCC) which are defined as
\begin{align}\label{def:measures}
& \text{Specificity} = \frac{\text{TN}}{\text{TN} + \text{FP}}, \qquad \text{ Sensitivity } = \frac{\text{TP}}{\text{TP} + \text{FN}}, \\
\nonumber& \text{MCC} = \frac{\text{TP}\times \text{TN} - \text{FP}\times \text{FN}}{\sqrt{(\text{TP} + \text{FP})(\text{TP}+\text{FN})(\text{TN}+ \text{FP})(\text{TN}+\text{FN})}}
\end{align}
with the TP, TN, FP, and FN representing the true positives (pCR), true negatives, false positives, and false negatives, respectively. For each of these three measures, the larger the value the better the classification performance.

\begin{figure}[ht!] \centering
\begin{center}%
\begin{tabular}
[c]{c}%
{\hspace{-0.55in} \includegraphics[trim = 0mm 0mm 0mm 0mm, clip, width=14cm, height=8cm]%
{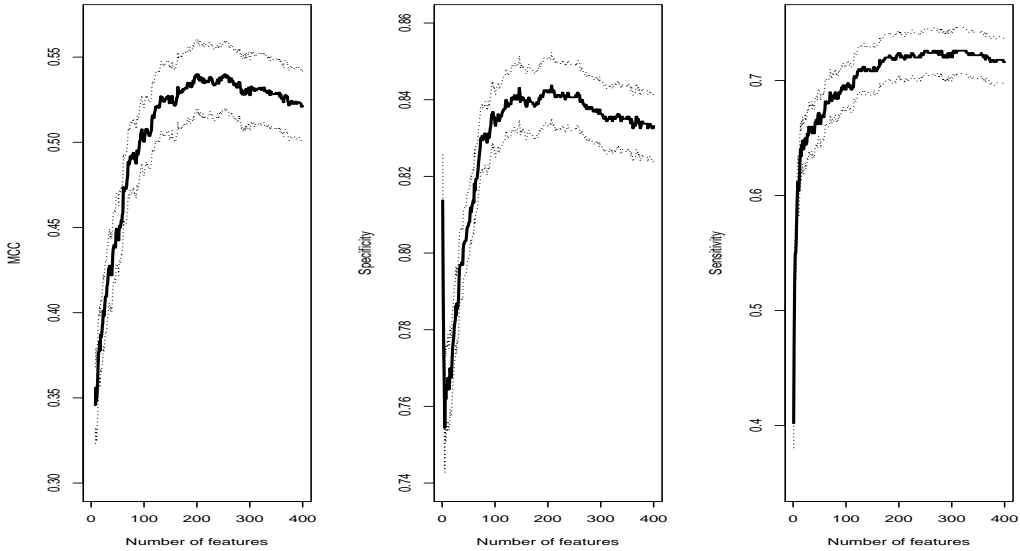}%
}%
\end{tabular}%
\caption{The classification results on breast cancer data with the $x$-axis representing the number of nonzero components in two-class mean difference estimate $T_{\tau}(\hmu)$.}
\label{Class}%
\end{center}%
\end{figure}%

As demonstrated in \cite{FF08}, feature selection can be crucial in high-dimensional classification since otherwise the noise accumulation caused by estimating a large number of parameters can dominate the signal and thus deteriorate the classification power. The same phenomenon is observed in our study here. When estimating the two class mean difference vector $\bmu_1 - \bmu_2$, we incorporate the feature selection component using the thresholded estimator $T_{\tau}(\hmu)$ defined similarly as in (\ref{092}). As the value of the threshold $\tau$ decreases, the number of nonzero components in $T_{\tau}(\hmu)$ varies from 1 to $p$. Figure \ref{Class} reports the three measures defined in (\ref{def:measures}) as functions of threshold $\tau$. To ease the presentation, we relabel the $x$-axis as the number of nonzero components in $T_{\tau}(\hmu)$. The solid curves are the mean values across 100 repetitions and the dotted curves around them are one standard error away from the mean curves pointwise.

The maximum value of MCC, which is 0.540, is achieved when 253 genes are used in LDA, and the corresponding standard error is 0.020. The values of MCC reported in \cite{CaiLiuLuo11} and \cite{FanFengWu09} are 0.506 and 0.402, respectively, with standard error both equal to 0.020, when using only the 110 most significant genes. Our results show that using a larger number of genes and taking into account their correlation structure have potential to improve the classification results. The specificity and sensitivity reported in \cite{FanFengWu09} are 0.794 (0.098) and 0.634 (0.220), respectively, with standard errors in parentheses, while the corresponding ones reported in \cite{CaiLiuLuo11} are 0.749 (0.005) and 0.806 (0.017), respectively. Comparing these results to Figure \ref{Class}, it is seen that we have much improved specificity and comparable sensitivity over a large region of the threshold level $\tau$.

\section{Discussions} \label{Sec6}
In this paper we have introduced a new method ISEE for efficient estimation of ultra-large Gaussian graphs. Thanks to its scalability, ISEE provides an effective way of uncovering large sparse graphs with big data. The suggested method is ideal for parallel and distributed computing and cloud computing and has been shown to enjoy appealing theoretical properties. Both computational and theoretical advantages of ISEE have been demonstrated with empirical studies. The ISEE can further scale up along with the use of the SIS or ISIS in \cite{FL08}; see Section \ref{SecA.supp.sis} of Supplementary Material for detailed descriptions of such an extension as well as its theoretical properties. 

It would be of interest to study several extensions of ISEE to different settings in future studies. For example, the idea of ISEE can be extended to the setting of large-scale multiple graphs comparison and estimation. Aided by ISEE, one can estimate each graph individually and then conduct multiple testing to detect the difference and similarity of these graphs. Another possible extension of ISEE is the estimation of large latent variable Gaussian graphical models, where only a subset of the nodes are observable in practice. It is also interesting to extend ISEE to the estimation of large nonparanormal graphical models, where the original graph for the $p$-variate random vector $\bx$ is non-normal, but under some unknown nonlinear transformation $\bff: \mathbb R^p \rightarrow \mathbb R^p$, $\bff(\bx)$ becomes a normal random vector.

As discussed in Section \ref{Sec3}, ISEE can be applied to such applications as 
dimension reduction; portfolio management; and multiple testing, feature screening, and simultaneous confidence intervals. It is interesting to investigate the performance of ISEE in these applications which demands future studies.

%
%
%
%
%

\appendix

\section{Proofs of some main results} \label{SecA}
We provide the proofs of Theorems \ref{Thm1}--\ref{Thm2} in this appendix. The proofs of Theorem \ref{Thm5} 
and Proposition \ref{Prop1} and additional technical details are included in the Supplementary Material.

\subsection{Proof of Theorem \ref{Thm1}} \label{A.1}
Throughout the proof we condition on the event $\mathcal{E}$ defined in (\ref{036}), with probability $1 - o\{p^{-(\delta - 2)}\}$ tending to one, on which the bounds (\ref{017})--(\ref{019}) hold simultaneously and uniformly over all nodes $j$ in the index sets $A_l$ with $1 \leq l \leq L$, and the entrywise $L_\infty$-norm bounds in (\ref{034}) hold uniformly as well. Observe that in view of (\ref{011})--(\ref{012}) and (\ref{014})--(\ref{038}), it is easy to see that the principal submatrix of the initial ISEE estimator $\ihOmeisee$, which is the sample covariance matrix $n^{-1} \widehat{\bX}\t \widehat{\bX}$, given by each index set $A_l$ is simply the matrix $\widehat{\Ome}_{A_l}$ given in (\ref{012}). Thus the uniform entrywise $L_\infty$-norm bound (\ref{023}) in Lemma \ref{Lem2} yields the bound
\begin{equation} \label{074}
\|n^{-1} \widehat{\bX}_{A_l}\t \widehat{\bX}_{A_l} - \Ome_{A_l}\|_\infty \leq O\left\{\max\left(K \lambda^2, \lambda\right)\right\}
\end{equation}
uniformly over the $L$ blocks of principal submatrices of $\ihOmeisee$ corresponding to the index sets $A_l$.

It remains to show that for each pair of index sets $(A_l, A_m)$ with $l \neq m$, we have
\begin{equation} \label{039}
\left\|n^{-1} \widehat{\bX}_{A_l}\t \widehat{\bX}_{A_m}\right\|_\infty = O\left\{\max\left(K \lambda^2, K^\alpha \lambda\right)\right\}.
\end{equation}
In light of (\ref{Xtilde_A})--(\ref{014}) and (\ref{024}), we have the following decomposition of the matrix
\begin{align}
\label{040}
\widehat{\bX}_{A_l} &= \bE_{A_l} \Ome_{A_l} + \bE_{A_l} \left(\widehat{\Ome}_{A_l} - \Ome_{A_l}\right) - \bX_{A_l^c} \left(\widehat{\bC}_{A_l} - \bC_{A_l}\right)  \widehat{\Ome}_{A_l} \\
\nonumber
& = \widetilde{\bX}_{A_l} + \bE_{A_l} \left(\widehat{\Ome}_{A_l} - \Ome_{A_l}\right) - \bX_{A_l^c} \left(\widehat{\bC}_{A_l} - \bC_{A_l}\right)  \widehat{\Ome}_{A_l},
\end{align}
where $\widehat{\bC}_{A_l} = (\hbbeta_{j,l})_{j \in A_l}$ denotes a $(p - |A_l|) \times |A_l|$ matrix of estimated regression coefficients. It follows from (\ref{040}) that
\begin{equation} \label{041}
n^{-1} \widehat{\bX}_{A_l}\t \widehat{\bX}_{A_m} = \bet_1 + \bet_2 + \bet_3 + \bet_4,
\end{equation}
where the first term is $\bet_1 = n^{-1} \widetilde{\bX}_{A_l}\t \widetilde{\bX}_{A_m}$, the second and third terms are $\bet_2 = n^{-1} \widetilde{\bX}_{A_l}\t [\bE_{A_m} (\widehat{\Ome}_{A_m} - \Ome_{A_m}) - \bX_{A_m^c} (\widehat{\bC}_{A_m} - \bC_{A_m})  \widehat{\Ome}_{A_m}]$ and  $\bet_3 = n^{-1} [\bE_{A_l} (\widehat{\Ome}_{A_l} - \Ome_{A_l}) - \bX_{A_l^c} (\widehat{\bC}_{A_l} - \bC_{A_l})  \widehat{\Ome}_{A_l}]\t \widetilde{\bX}_{A_m}$, and the last term is $\bet_4 = n^{-1} [\bE_{A_l} (\widehat{\Ome}_{A_l} - \Ome_{A_l}) - \bX_{A_l^c} (\widehat{\bC}_{A_l} - \bC_{A_l})  \widehat{\Ome}_{A_l}]\t [\bE_{A_m} (\widehat{\Ome}_{A_m} - \Ome_{A_m}) - \bX_{A_m^c} (\widehat{\bC}_{A_m} - \bC_{A_m})  \widehat{\Ome}_{A_m}]$. We will analyze these four terms separately.

\medskip

\textit{Part 1}. We start with the second and third terms $\bet_2$ and $\bet_3$. Since $\widetilde{\bX}_{A_l} = \bE_{A_l} \Ome_{A_l}$, we can rewrite $\bet_2$ as
\begin{align} \label{042}
\bet_2 & = n^{-1} \Ome_{A_l}\t \bE_{A_l}\t \left[\bE_{A_m} \left(\widehat{\Ome}_{A_m} - \Ome_{A_m}\right) - \bX_{A_m^c} \left(\widehat{\bC}_{A_m} - \bC_{A_m}\right) \widehat{\Ome}_{A_m}\right] \\
\nonumber
& = \bD_1 - \bD_2,
\end{align}
where $\bD_1 = \Ome_{A_l}\t (n^{-1} \bE_{A_l}\t \bE_{A_m}) (\widehat{\Ome}_{A_m} - \Ome_{A_m})$ and $\bD_2 = \Ome_{A_l}\t (n^{-1} \bX_{A_m^c}\t \bE_{A_l})\t \\ \cdot (\widehat{\bC}_{A_m} - \bC_{A_m}) \widehat{\Ome}_{A_m}$. Note that the error matrices $\bE_{A_l}$ and $\bE_{A_m}$ with $l \neq m$ are independent of each other and thus the mean of the random matrix $n^{-1} \bE_{A_l}\t \bE_{A_m}$ is $\bzero$. So the same concentration bound as in (\ref{028}) applies with $\bxi_1$ replaced by $n^{-1} \bE_{A_l}\t \bE_{A_m}$. Taking $t = [(\delta + 1) (\log p)/(cn)]^{1/2}$ in (\ref{028}) and applying Bonferroni's inequality over $1 \leq l \neq m \leq L$ lead to
\begin{equation} \label{043}
P(\mathcal{E}_3) \geq 1 - p^2 \cdot O(e^{-c n t^2}) = 1 - O\left\{p^{-(\delta - 1)}\right\} = 1 - o\left\{p^{-(\delta - 2)}\right\},
\end{equation}
where the event $\mathcal{E}_3$ is defined as
\begin{equation} \label{044}
\mathcal{E}_3 = \left\{\max_{1 \leq l \neq m \leq L} \left\|n^{-1} \bE_{A_l}\t \bE_{A_m}\right\|_\infty \leq t = O(\lambda)\right\}.
\end{equation}
From now on, we condition on the event $\mathcal{E} \cap \mathcal{E}_3$, which has the same asymptotic probability bound as $\mathcal{E}$ in view of (\ref{035}) and (\ref{043}). As shown in the proof of Lemma \ref{Lem2}, it holds that $\|\Ome_{A_l}\|_\infty = O(1)$ and $\|\widehat{\Ome}_{A_l}\|_\infty = O(1)$. Thus on the event $\mathcal{E} \cap \mathcal{E}_3$, we have
\begin{equation} \label{045}
\|\bD_1\|_\infty = O\left\{\lambda \max\left(K \lambda^2, \lambda\right)\right\}
\end{equation}
in view of (\ref{044}) and (\ref{023}).

For the second term $\bD_2$ in (\ref{042}), consider the $|A_l| \times |A_m|$ matrix
\begin{equation} \label{069}
\bF = (n^{-1} \bX_{A_m^c}\t \bE_{A_l})\t (\widehat{\bC}_{A_m} - \bC_{A_m}) = \bF_ 1 + \bF_2,
\end{equation}
where $\bF_1$ and $\bF_2$ are defined through matrix multiplication by taking the rows of $n^{-1} \bX_{A_m^c}\t \bE_{A_l}$ and $\widehat{\bC}_{A_m} - \bC_{A_m}$ from nodes in index sets $A_m^c \cap A_l^c$ and $A_l$, respectively. In view of (\ref{019}) and (\ref{017}), we have
\begin{equation} \label{061}
\|\bF_1\|_\infty \leq O(\lambda) \cdot O(K \lambda) = O(K \lambda^2).
\end{equation}
Denote by $\bF_3$ the $|A_l| \times |A_m|$ submatrix of $\widehat{\bC}_{A_m} - \bC_{A_m}$ given by rows corresponding to nodes in $A_l$. By Lemma \ref{Lem3}, Theorem 3 of \cite{YZ10} applies to show that
\begin{equation} \label{062}
P(\mathcal{E}_4) = 1 - o\left\{p^{-(\delta -2)}\right\},
\end{equation}
where $\mathcal{E}_4 = \{\max_{j \in A_l, 1 \leq l \leq L}\|\hbbeta_{j,l} - \bbeta_{j,l}\|_\infty = O(K^\alpha \lambda)\}$. In view of (\ref{004})--(\ref{005}), using similar arguments to those for proving (\ref{029}) with $t$ chosen to be $[\delta (\log p)/(cn)]^{1/2}$ leads to
\begin{equation} \label{063}
P(\mathcal{E}_5) = 1 - o\left\{p^{-(\delta -2)}\right\},
\end{equation}
where $\mathcal{E}_5 = \{\max_{1 \leq l \leq L} \|n^{-1} \bX_{A_l}\t \bE_{A_l} - \Ome_{A_l}^{-1}\|_\infty \leq O(\lambda)\}$.

Hereafter we condition on the event $\mathcal{E} \cap (\cap_{3 \leq i \leq 5} \mathcal{E}_i)$, which has the same asymptotic probability bound as $\mathcal{E}$ in view of (\ref{035}), (\ref{043}), and (\ref{062})--(\ref{063}). On this new event, it follows from (\ref{062})--(\ref{063}) and the fact of $\|\Ome_{A_l}^{-1}\|_\infty = O(1)$ that
\begin{equation} \label{064}
\|\bF_2\|_\infty = \left\|\left(n^{-1} \bX_{A_l}\t \bE_{A_l}\right)\t \bF_3\right\|_\infty \leq O(1) \cdot O(K^\alpha \lambda) = O(K^\alpha \lambda).
\end{equation}
Combining (\ref{069})--(\ref{061}) and (\ref{064}) together with the facts of $\|\Ome_{A_l}\|_\infty = O(1)$ and $\|\widehat{\Ome}_{A_m}\|_\infty = O(1)$ gives
\begin{equation} \label{065}
\|\bD_2\|_\infty = O\left\{\max\left(K \lambda^2, K^\alpha \lambda\right)\right\}.
\end{equation}
Since $\bet_3\t$ shares the same form as $\bet_2$, putting (\ref{042}), (\ref{045}), and (\ref{065}) together yields
\begin{align} \label{066}
P& \left\{ \max_{1 \leq l \neq m \leq L}\max(\|\bet_2\|_\infty, \|\bet_3\|_\infty) \leq O\left\{\max\left(K \lambda^2, K^\alpha \lambda\right)\right\}\right\} \\
\nonumber
& = 1 - o\left\{p^{-(\delta -2)}\right\}.
\end{align}

\medskip

\textit{Part 2}. We next consider the fourth term $\bet_4$. Let us decompose it into four terms as
\begin{equation} \label{067}
\bet_4 = \bG_1 - \bG_2 - \bG_3 + \bG_4,
\end{equation}
where the first term is $\bG_1 = n^{-1} (\widehat{\Ome}_{A_l} - \Ome_{A_l}) \bE_{A_l}\t \bE_{A_m} (\widehat{\Ome}_{A_m} - \Ome_{A_m})$, the second and third terms are $\bG_2 = n^{-1} (\widehat{\Ome}_{A_l} - \Ome_{A_l}) \bE_{A_l}\t \bX_{A_m^c} (\widehat{\bC}_{A_m} - \bC_{A_m})  \widehat{\Ome}_{A_m}$ and $\bG_3 = n^{-1} \widehat{\Ome}_{A_l} (\widehat{\bC}_{A_l} - \bC_{A_l})\t \bX_{A_l^c}\t \bE_{A_m} (\widehat{\Ome}_{A_m} - \Ome_{A_m})$, and the last term is $\bG_4 = n^{-1} \widehat{\Ome}_{A_l} (\widehat{\bC}_{A_l} - \bC_{A_l})\t \bX_{A_l^c}\t \bX_{A_m^c} (\widehat{\bC}_{A_m} - \bC_{A_m})  \widehat{\Ome}_{A_m}$. In view of (\ref{036}) and (\ref{044}), we see that on the event $\mathcal{E} \cap \mathcal{E}_3$, it holds that
\begin{align} \label{068}
\|\bG_1\|_\infty & \leq O\left\{\max\left(K \lambda^2, \lambda\right)\right\}  \cdot O(\lambda) \cdot O\left\{\max\left(K \lambda^2, \lambda\right)\right\} \\
\nonumber
& = O\left\{\lambda \left[\max\left(K \lambda^2, \lambda\right)\right]^2\right\}.
\end{align}
Note that $\bG_2 = (\widehat{\Ome}_{A_l} - \Ome_{A_l}) \bF \widehat{\Ome}_{A_m}$ in light of (\ref{069}), and $\bG_3\t$ shares the same form as $\bG_2$. Thus on the event $\mathcal{E} \cap (\cap_{3 \leq i \leq 5} \mathcal{E}_i)$, combining (\ref{023}), (\ref{069})--(\ref{061}), and (\ref{064}) along with the fact of $\|\widehat{\Ome}_{A_m}\|_\infty = O(1)$ leads to
\begin{equation} \label{070}
\max\left(\|\bG_2\|_\infty, \|\bG_3\|_\infty\right) \leq O\left\{\max\left(K \lambda^2, \lambda\right) \cdot \max\left(K \lambda^2, K^\alpha \lambda\right)\right\}.
\end{equation}

For the last term $\bG_4$, observe that an application of the Cauchy-Schwarz inequality together with bound (\ref{018}) entails
\[ \|n^{-1} (\widehat{\bC}_{A_l} - \bC_{A_l})\t \bX_{A_l^c}\t \bX_{A_m^c} (\widehat{\bC}_{A_m} - \bC_{A_m})\|_\infty \leq O(K \lambda^2). \]
Since $\|\widehat{\Ome}_{A_l}\|_\infty = O(1)$, it follows from the above bound that
\begin{equation} \label{071}
\|\bG_4\|_\infty \leq O(K \lambda^2).
\end{equation}
Thus combining (\ref{067})--(\ref{071}) results in
\begin{equation} \label{072}
P \left\{ \max_{1 \leq l \neq m \leq L} \|\bet_4\|_\infty \leq O\left\{\max\left(K \lambda^2, \lambda\right)\right\}\right\} = 1 - o\left\{p^{-(\delta -2)}\right\}.
\end{equation}

\medskip

\textit{Part 3}. We finally consider the first term $\bet_1$. In view of (\ref{007})--(\ref{008}), $n^{-1} \widetilde{\bX}\t \widetilde{\bX}$ is the oracle sample covariance matrix estimator for the precision matrix $\Ome$. Since $\widetilde{\bx}$ defined in (\ref{006}) is a $p$-variate Gaussian random vector, applying similar arguments to those for proving (\ref{029}) with $t$ chosen to be $[(\delta + 1) (\log p)/(cn)]^{1/2}$ leads to
\begin{equation} \label{073}
P\left\{\|n^{-1} \widetilde{\bX}\t \widetilde{\bX} - \Ome\|_\infty \leq O(\lambda)\right\} = 1 - o\left\{p^{-(\delta -2)}\right\},
\end{equation}
which provides a uniform bound on $\bet_1 = n^{-1} \widetilde{\bX}_{A_l}\t \widetilde{\bX}_{A_m}$.

Therefore, combining (\ref{041}), (\ref{066}), and (\ref{072})--(\ref{073}) gives bound (\ref{039}) uniformly over all pairs of index sets $(A_l, A_m)$ with $l \neq m$. Observe that the order in (\ref{039}) is in fact $O(K^\alpha \lambda)$ since the rate of convergence $O(K^{\alpha}\lambda)$ dominates that of $O(K\lambda^2)$ in light of the assumptions of $K^{1+\alpha}\lambda = o(1)$ and $\alpha \geq 0$. Then in view of (\ref{074}), the proof of Theorem \ref{Thm1} concludes by noticing that all these uniform bounds hold simultaneously with significant probability $1 - o\{p^{-(\delta - 2)}\}$.

\subsection{Proof of Theorem \ref{Thm2}} \label{A.2}
Since $C$ in $\omega^*_0 = C  K^\alpha \lambda$ is some sufficiently large positive constant, Theorem \ref{Thm1} entails that with probability $1 - o\{p^{-(\delta - 2)}\}$, it holds that
\begin{equation} \label{095}
\left\|\ihOmeisee - \Ome\right\|_\infty < c \omega^*_0,
\end{equation}
where $c < 1/2$ is some positive constant. Thus in view of the assumption that $\omega_0 = \min\{|\omega_{jk}|: (j, k) \in \supp(\Ome)\} \geq \omega^*_0$, by (\ref{095}) we have $\supp(\ghOmeisee) \subset \supp(\Ome)$ when $\tau \geq c \omega^*_0$, and $\supp(\Ome) \subset \supp(\ghOmeisee)$ when $\tau \leq \omega_0 - c \omega^*_0$. This shows that $\supp(\ghOmeisee) = \supp(\Ome)$ for any $\tau \in [c \omega^*_0, \omega_0 - c \omega^*_0]$, which proves part a of Theorem \ref{Thm2}.

For part b, we first make an important claim that the results of Theorem \ref{Thm1} hold for the initial ISEE estimator $\ihOmeisee$ as defined in (\ref{038}) but based on a subsample of $n_0$ rows of the estimated oracle empirical matrix $\widetilde{\bX}$, where $n_0/n$ is bounded away from $0$. This claim follows from the same arguments as in the proof of Theorem \ref{Thm1}, by noting that bounds (\ref{018})--(\ref{019}) hold when the subsample is used since $n_0$ is of the same order as $n$. Observe that both $n_1$ and $n_2$ are of the same order as $n$ by the assumption that $n_1/n_2$ is bounded away from $0$ and $\infty$. Thus with probability $1 - o\{p^{-(\delta - 2)}\}$, the bound (\ref{095}) also applies to both estimators $\ihOmeisee^{1, \nu}$ and $\ihOmeisee^{2, \nu}$, that is,
\begin{equation} \label{098}
\left\|\ihOmeisee^{i, \nu} - \Ome\right\|_\infty < c \omega^*_0
\end{equation}
for $i = 1, 2$.

As in \cite{BL08b}, without loss of generality we work with the case of $N = 1$ in (\ref{094}). In view of the proof for part a, to prove the sure screening property $\supp(\Ome) \subset \supp(\ghOmeisee)$ it suffices to show that with probability $1 - o\{p^{-(\delta - 2)}\}$, the threshold $\tau$ chosen by the cross-validation is bounded above by $\tau_0 = \omega_0 - c \omega^*_0$. Here we use the convention that the smallest $\tau$ is preferred when the minimizer of $R(\tau)$ is not unique. To this end, we need only to show that $R(\tau) \geq R(\tau_0)$ whenever $\tau \geq \tau_0$.

Note that when $\tau = \tau_0$, we have $\supp\{T_\tau(\ihOmeisee)\} = \supp(\Ome)$ by part a, and $\supp\{T_\tau(\ihOmeisee^{i, \nu})\} = \supp(\Ome)$ for $i = 1, 2$ in light of (\ref{098}). Thus when $\tau$ increases from $\tau_0$, the two matrices $T_\tau(\ihOmeisee^{1, \nu})$ and $T_{\tau_0}(\ihOmeisee^{1, \nu})$ can differ only over entries in $\supp(\Ome)$. Assume that $M$ nonzero entries of $T_{\tau_0}(\ihOmeisee^{1, \nu})$ become zero in $T_\tau(\ihOmeisee^{1, \nu})$. Then by some simple algebra, it follows from (\ref{098}) and the assumption of $\omega_0 = \min\{|\omega_{jk}|: (j, k) \in \supp(\Ome)\} \geq \omega^*_0$ that
\begin{align} \label{099}
R(\tau) - R(\tau_0) & \geq M \left[(\omega_0 - c \omega^*_0)^2 - (2 c \omega^*_0)^2\right] \\
\nonumber
& \geq M (1 - 3 c) (1 + c) (\omega^*)_0^2 \geq 0,
\end{align}
as long as we choose $c \leq 1/3$ in part a. This complets the proof of part b of Theorem \ref{Thm2}.

Finally for part c, note that $\ghOmeisee$ from either of parts a and b satisfies that $\supp(\Ome) \subset \supp(\ghOmeisee)$. Using the same arguments as in the proof of Lemma \ref{Lem2} with $t$ chosen as $[(\delta + 1) (\log p)/(cn)]^{1/2}$ in (\ref{028}), we can show that with probability $1 - o\{p^{-(\delta - 2)}\}$, it holds uniformly over all pairs of nodes $A = \{j, k\}$ that $\|\widehat{\Ome}_A - \Ome_{A,A}\|_\infty = O\{\max(K \lambda^2, \lambda)\}$. This result along with $\supp(\Ome) \subset \supp(\ghOmeisee)$ yields the desired bound (\ref{093}) in part c, by noting that the order $O(\lambda)$ dominates $O(K \lambda^2)$ in view of the assumptions of $K^{1+\alpha}\lambda = o(1)$ and $\alpha \geq 0$. We conclude the proof of part c of Theorem \ref{Thm2} by showing the sure screening property which can be exploited to reduce the computational cost of the refinement step for estimating the link strength. When the ISEE estimator with refinement $\hOmeisee$ updates the $(j, k)$-entry of $\ghOmeisee$, two univariate linear regression models as defined in (\ref{009}) with $A = \{j, k\}$ are considered for nodes $j$ and $k$, respectively. In light of $\bC_A = -\Ome_{A^c, A} \Ome_{A,A}^{-1}$ in model (\ref{005}), it is easy to see that
\begin{equation} \label{096}
\supp(\bbeta_j), \supp(\bbeta_k) \subset \left\{m \in A^c: |\omega_{jm}| \text{ or } |\omega_{km}| \neq 0 \right\}.
\end{equation}
Denote by $\widehat{M}_{jk} = \{m \in A^c: |\widehat{\omega}_{jm}| \text{ or } |\widehat{\omega}_{km}| \neq 0\}$, where $\ghOmeisee = (\widehat{\omega}_{jk})$. Thus by (\ref{096}) and $\supp(\Ome) \subset \supp(\ghOmeisee)$, with probability $1 - o\{p^{-(\delta - 2)}\}$ it holds uniformly over all pairs of nodes $(j, k)$ that
\begin{equation} \label{097}
\supp(\bbeta_j), \supp(\bbeta_k) \subset \widehat{M}_{jk},
\end{equation}
which gives the desired sure screening property for fitting model (\ref{009}).

\begin{supplement}[id=ISEE]
\stitle{Supplementary material to ``Innovated Scalable Efficient Estimation in Ultra-Large Gaussian Graphical Models''}
  \slink[doi]{10.1214/00-AOSXXXXSUPP}
  \sdatatype{.pdf}
  \sdescription{Due to space constraints, the proofs of Theorem \ref{Thm5} 
and Proposition \ref{Prop1} and additional technical details are provided in the Supplementary Material \citep{FanLv2015}.}
\end{supplement}

\bigskip
{\scriptsize
\begin{tabular}{ll}
\hspace{2.14in} & \textsc{Data Sciences and Operations Department}\\
& \textsc{Marshall School of Business}\\
& \textsc{University of Southern California}\\
& \textsc{Los Angeles, CA 90089}\\
& \textsc{USA}\\
& \textsc{\printead{e1}}\\
& \textsc{\phantom{E-mail:\ }\printead*{e2}}
\end{tabular}}




\newpage
\title{Supplementary Material to ``Innovated Scalable Efficient Estimation in Ultra-Large Gaussian Graphical Models''}

\begin{aug}
\author{\fnms{Yingying} \snm{Fan}\ead[label=e1]{fanyingy@marshall.usc.edu}}
\and
\author{\fnms{Jinchi} \snm{Lv}\ead[label=e2]{jinchilv@marshall.usc.edu}}

\runauthor{Y. Fan and J. Lv}

\affiliation{University of Southern California}
\end{aug}

\bigskip

This Supplementary Material contains the proofs of Theorem \ref{Thm5}, Proposition \ref{Prop1}, and additional technical details, as well as an extension of ISEE by incorporating the idea of feature screening.

\smallskip

\appendix
\setcounter{page}{1}
\setcounter{section}{1}
\renewcommand{\theequation}{A.\arabic{equation}}
\setcounter{equation}{0}

\section{Ultra-large graph screening} \label{SecA.supp.sis}

\subsection{SIS-assisted ISEE} \label{Sec2.5}

When the scale of the number of nodes $p$ is ultra large, we can exploit the sure independence screening (SIS) in \cite{FL08} to reduce the computational cost for each scaled Lasso regression. 
For each node $j$ in the index set $A_l$ with $1 \leq l \leq L$, the SIS ranks the components of the vector
\begin{equation} \label{075}
\bw = (w_k)_{k \in A_l^c} = \bX_{A_l^c}\t \bX_j
\end{equation}
obtained by componentwise regression and for any given $\zeta \in (0, 1)$, defines a submodel
\begin{equation} \label{076}
\mathcal{M}_{jl,\zeta} = \left\{k \in A_l^c: \text{ $|w_k|$ is among the first $[\zeta n]$ largest of all}\right\},
\end{equation}
where $[\zeta n]$ denotes the integer part of $\zeta n$. Here for simplicity, each node random variable $X_j$ is assumed to have standard deviation one as in \cite{FL08}.

Following \cite{FL08}, based on the reduced model $\mathcal{M}_{jl,\zeta}$ obtained by the SIS one can construct the SIS-SLasso estimator $\hbbeta^*_{j,l}$, which is the scaled Lasso estimator $\hbbeta_{j,l}$ as defined in (\ref{010}) with zero components outside the index set $\mathcal{M}_{jl,\zeta}$ for $\bbeta$. Similarly as in (\ref{038}), we define the initial ISEE estimator $\ihOmeisee^*$ as the sample covariance matrix
\begin{equation} \label{091}
\ihOmeisee^* = n^{-1} (\widehat{\bX}^*)\t \widehat{\bX}^*,
\end{equation}
where the estimator $\widehat{\bX}^*$ for the oracle empirical matrix $\widetilde{\bX}$ is constructed as in (\ref{014}) using the SIS-SLasso estimator $\hbbeta^*_{j,l}$. Then we can construct the ISEE estimator for the graph $\ghOmeisee$ and the ISEE estimator with refinement $\hOmeisee$ based on the SIS-assisted initial ISEE estimator $\ihOmeisee^*$ in (\ref{091}) as described in Section \ref{Sec2.3}. Similarly the iterative SIS (ISIS) in \cite{FL08} can also be applied to improve over the SIS in ultra-large scale problems.

\subsection{Technical conditions}

\begin{condition} \label{Cond3}
It holds that $p > n$ and $\log p = O(n^\gamma)$ for some constant $0 < \gamma < 1 - 2 \kappa$ with $\kappa$ defined in Condition \ref{Cond4}.
\end{condition}

\begin{condition} \label{Cond4}
There exist some constants $0 \leq \kappa < 1/2$ and $c_1, c_2, c_3 > 0$ such that for each $j \in A_l$ with $1 \leq l \leq L$, the support of the regression coefficient vector $\bbeta_{j,l} = (\beta_{jlk})_{k \in A_l^c}$ in (\ref{009}) admits a decomposition $\supp(\bbeta_{j,l}) = S_{jl0} \cup S_{jl1}$, where for each $k \in S_{jl0}$, $|\beta_{jlk}| \geq c_1 n^{-\kappa}$ and $|\cov(\beta_{jlk}^{-1} X_j, X_k)| \geq c_2$, and for each $k \in A_l^c$, $|\cov(\sum_{m \in S_{jl1}} \beta_{jlm} X_m, X_k)| \leq c_3 \lambda$. Moreover, it holds that
\begin{equation} \label{083}
\max_{j \in A_l, \, 1 \leq l \leq L} \max\Big\{\sum_{m \in S_{jl1}} |\beta_{jlm}|, \lambda^{-1} \sum_{m \in S_{jl1}} \beta_{jlm}^2\Big\} = O(K \lambda).
\end{equation}
\end{condition}

Conditions \ref{Cond3} and \ref{Cond4} are additional assumptions that facilitate the analysis of the SIS-assisted ISEE approach and ensure the sure screening property of the SIS procedure as in \cite{FL08}. In particular, Condition \ref{Cond3} allows the dimensionality $p$ to increase exponentially with sample size $n$. Condition \ref{Cond4} is imposed to ensure that the SIS-assisted ISEE estimate can enjoy nice asymptotic properties.

\subsection{Theoretical properties}

As introduced in Section \ref{Sec2.5}, to reduce the computational cost we can apply ISEE along with SIS or ISIS in the initial step for ultra-large graph screening. The computational cost can be further reduced if we also apply SIS or ISIS in the refinement step of estimating the link strength. 
In the refinement step, 
for each identified link $(j, k)$ we can fit model (\ref{009}) instead on the union of the supports of the $j$th and $k$th rows of $\ghOmeisee$, with nodes $j$ and $k$ excluded; see (\ref{097}) in the proof of Theorem \ref{Thm2} for more details.

The following two theorems characterize the performance of the SIS-assisted ISEE estimators in both the initial step and the refinement step.


\begin{theorem} \label{Thm3}
Assume that the conditions of Theorem \ref{Thm1} and Conditions \ref{Cond3}--\ref{Cond4} hold and $\zeta$ in \eqref{076} is at least of order $n^{-\gamma_0}$ with some constant $0 < \gamma_0 < 1 - 2 \kappa$. Then the SIS-assisted initial  ISEE estimator $\ihOmeisee^*$ in (\ref{091}) satisfies the same properties as in Theorem \ref{Thm1}.
\end{theorem}


\begin{theorem} \label{Thm4}
Under the conditions of Theorems \ref{Thm2} and \ref{Thm3}, the ISEE estimators $\ghOmeisee$ and $\hOmeisee$ based on 
$\ihOmeisee^*$ in (\ref{091}) satisfy the same properties as in Theorem \ref{Thm2}.
\end{theorem}

%

\section{Proofs of additional main results} \label{SecA.supp}

\subsection{Proof of Theorem \ref{Thm5}} \label{A.5}
By (\ref{038}), (\ref{074}), and the definition of the bias corrected initial ISEE estimator $\cihOmeisee$ in (\ref{bcisee01}) and (\ref{bcisee02}), it suffices to consider the off-block-diagonal entries of the initial ISEE estimator $\ihOmeisee$, that is, the submatrices $(\ihOmeisee)_{A_l, A_m}$ with $1 \leq l \neq m \leq L$. The bias of the initial ISEE estimator $\ihOmeisee$ comes from these entries. Note that for each $l \neq m$, $(\ihOmeisee)_{A_l, A_m}$ admits the representation in (\ref{041}). By (\ref{072}) and (\ref{073}), we see that the aforementioned bias is incurred by the second and third terms $\bet_2$ and $\bet_3$ in (\ref{041}).

Due to the symmetry, we focus only on the term $\bet_2$. Examining Part 1 of the proof of Theorem \ref{Thm1}, we see that the bias in the term $\bet_2$ is caused only by the additive component
\begin{equation} \label{107}
\widetilde{\bF}_2 = -\Ome_{A_l}\t \bF_2 \widehat{\Ome}_{A_m},
\end{equation}
where $\bF_2$ defined in (\ref{069}) is given by $(n^{-1} \bX_{A_l}\t \bE_{A_l})\t (\widehat{\bC}_{A_m}^{A_l} - \bC_{A_m}^{A_l})$, and $\widehat{\bC}_{A_m}^ {A_l}$ and $\bC_{A_m}^ {A_l}$ denote submatrices of $\widehat{\bC}_{A_m}$ and $\bC_{A_m}$ consisting of rows with indices in $A_l$, respectively. We now add a bias correction term $\widehat{\bC}_{A_m}^ {A_l} \widehat{\Ome}_{A_m}$ as specified in (\ref{bcisee02}) to $(\ihOmeisee)_{A_l, A_m}$, and subsequently to $\widetilde{\bF}_2$ given in (\ref{107}). Let us consider the resulting new term
\begin{equation} \label{108}
\widetilde{\bF}^*_2 = \widetilde{\bF}_2 + \widehat{\bC}_{A_m}^ {A_l} \widehat{\Ome}_{A_m} = \bF_4 + \bF_5 + \bC_{A_m}^{A_l} \Ome_{A_m},
\end{equation}
where $\bF_4 = -[\Ome_{A_l}\t (n^{-1} \bX_{A_l}\t \bE_{A_l})\t -I_{|A_l|}] (\widehat{\bC}_{A_m}^{A_l} - \bC_{A_m}^{A_l}) \widehat{\Ome}_{A_m}$ and $\bF_5 = \bC_{A_m}^{A_l} (\widehat{\Ome}_{A_m} - \Ome_{A_m})$. We study these two terms $\bF_4$ and $\bF_5$ separately.

As in Part 1 of the proof of Theorem \ref{Thm1}, we condition on the event $\mathcal{E} \cap (\cap_{3 \leq i \leq 5} \mathcal{E}_i)$ hereafter. Note that $\widehat{\bC}_{A_m}^{A_l} - \bC_{A_m}^{A_l}$ is exactly the matrix $\bF_3$ introduced therein. In light of the definitions of $\mathcal{E}$, $\mathcal{E}_4$, and $\mathcal{E}_5$ in (\ref{036}) and (\ref{062})--(\ref{063}), by the facts of $\|\Ome_{A_l}\|_\infty = O(1)$ and $\widehat{\Ome}_{A_m} = O(1)$ it holds uniformly over $1 \leq l \neq m \leq L$ that
\begin{align} \label{109}
\|\bF_4\|_\infty & \leq O(1) \|n^{-1} \bX_{A_l}\t \bE_{A_l} - \Ome_{A_l}^{-1}\|_\infty \|\bF_3\|_\infty O(1) \\ \nonumber
& \leq O(\lambda) \cdot O(K^\alpha \lambda) = O(K^\alpha \lambda^2).
\end{align}
Using similar arguments to those in the proof of Lemma \ref{Lem2}, we can show that $\|\bC_{A_m}^{A_l}\|_\infty = \|-\Ome_{A_l,A_m} \Ome_{A_m}^{-1}\|_\infty = O(1)$, which along with (\ref{036}) entails
\begin{equation} \label{110}
\|\bF_5\|_\infty = O\left\{\max(K \lambda^2, \lambda)\right\}.
\end{equation}
Since $\alpha \leq 1/2$ by Condition \ref{Cond2}, it follows from (\ref{109}) and (\ref{110}) that
\begin{equation} \label{111}
\|\bF_4 + \bF_5\|_\infty \leq O\left\{\max(K \lambda^2, \lambda)\right\}.
\end{equation}

Observe that $\bC_{A_m}^{A_l} \Ome_{A_m} = -\Ome_{A_l,A_m} \Ome_{A_m}^{-1} \Ome_{A_m} = -\Ome_{A_l,A_m}$. Therefore, combining (\ref{108}) and (\ref{111}) proves the desired bound for the bias corrected initial ISEE estimator $\cihOmeisee$ with off-block-diagonal entries
\[
\big(\cihOmeisee\big)_{A_l, A_m} = -\big[\big(\ihOmeisee\big)_{A_l, A_m} + \widehat{\bC}_{A_l}^ {A_m} \widehat{\Ome}_{A_l} + \widehat{\bC}_{A_m}^ {A_l} \widehat{\Ome}_{A_m}\big];
\]
that is, with the same probability bound as in Theorem \ref{Thm1} it holds that
\[
\left\|\cihOmeisee - \Ome\right\|_\infty = O\left\{\max(K \lambda^2, \lambda)\right\},
\]
which order is in fact $O(\lambda)$ as explained in the proof of Theorem \ref{Thm2}.

The second part of Theorem \ref{Thm5}, which is graph recovery consistency of the bias corrected initial ISEE estimator $\cihOmeisee$, can be proved using similar arguments to those in the proof for part a of Theorem \ref{Thm2}, by noting that $\ghOmeisee = T_\tau(\cihOmeisee)$ and $\omega^*_0 = C \lambda$ with $C > 0$ some sufficiently large constant.

\subsection{Proof of Theorem \ref{Thm3}} \label{A.3}
We first show that the two events $\mathcal{H}_1$ and $\mathcal{H}_2$ defined as
\begin{equation} \label{077}
\mathcal{H}_1 = \bigcap\nolimits_{j \in A_l, 1 \leq l \leq L} \left\{S_{jl0} \subset \mathcal{M}_{jl,\zeta}\right\}
\end{equation}
and
\begin{equation} \label{078}
\mathcal{H}_2 = \left\{\max\nolimits_{j \in A_l, 1 \leq l \leq L} \left\|n^{-1} \bX_{A_l^c}\t \sum\nolimits_{m \in S_{jl1}} \beta_{jlm} \bX_m\right\|_\infty \leq O(\lambda)\right\}
\end{equation}
have large probabilities. The event $\mathcal{H}_1$ in (\ref{077}) characterizes the sure screening property of the SIS associated with the sets of indices $S_{jl0}$. It is easy to check that Conditions 1--4 in \cite{FL08} 
are entailed by our Conditions \ref{Cond1} and \ref{Cond3}--\ref{Cond4}, with $\mathcal{M}_*$ replaced by $S_{jl0}$. In particular, they verified the property C (a concentration property) for Gaussian distributions.

A key observation is that the proof of Theorem 1 in \cite{FL08}
applies equally well to the case where the set of desired effects $S_{jl0}$ plays the role of $\mathcal{M}_*$ and the set of additional effects $S_{jl1} = \supp(\bbeta_{j,l}) \setminus S_{jl0}$ may not be empty. Thus an application of the same arguments leads to a similar conclusion to that in Theorem 1 of
\cite{FL08}; that is, for $\zeta$ at least in the order of $n^{-\gamma_0}$ with some positive constant $\gamma_0 < 1 - 2 \kappa$, we have
\begin{equation} \label{079}
P\left\{S_{jl0} \subset \mathcal{M}_{jl,\zeta}\right\} = 1 - O\left\{\exp[-C n^{1 - 2 \kappa}/(\log n)]\right\},
\end{equation}
where $C$ is some positive constant. Since $\log p = O(n^\gamma)$ with constant $0 < \gamma < 1 - 2 \kappa$ by Condition \ref{Cond3}, we see immediately from (\ref{079}) and Bonferroni's inequality over all nodes $j$ in the index sets $A_l$ that
\begin{equation} \label{080}
P(\mathcal{H}_1) \geq 1 - p \cdot o\left\{p^{-(\delta -1)}\right\} = 1 - o\left\{p^{-(\delta -2)}\right\}.
\end{equation}

Note that for each $k \in A_l^c$, the expectation of $n^{-1} \bX_k\t \sum\nolimits_{m \in S_{jl1}} \beta_{jlm} \bX_m$ is equal to $\cov(\sum_{m \in S_{jl1}} \beta_{jlm} X_m, X_k)$. Thus in view of the assumption of $\max_{k \in A_l^c} |\cov(\sum_{m \in S_{jl1}} \beta_{jlm} X_m, X_k)| \leq c_3 \lambda$ by Condition \ref{Cond4}, using similar arguments to those for proving (\ref{029}) with $t$ chosen to be $[(\delta + 1) (\log p)/(cn)]^{1/2}$ leads to
\begin{equation} \label{081}
P(\mathcal{H}_2) \geq 1 - p (p - 1) \cdot O\left\{p^{-(\delta + 1)}\right\} = 1 - o\left\{p^{-(\delta -2)}\right\}.
\end{equation}
Combining (\ref{080}) and (\ref{081}) yields the desired probability bound
\begin{equation} \label{082}
P(\mathcal{H}_1 \cap \mathcal{H}_2) \geq 1 - o\left\{p^{-(\delta -2)}\right\}.
\end{equation}

From now on we condition on the event $\mathcal{H}_1 \cap \mathcal{H}_2$. On this event, for each node $j$ in the index set $A_l$, the submodel $\mathcal{M}_{jl,\zeta}$ given by the SIS contains the set of desired effects $S_{jl0}$. In light of (\ref{078}), we can treat the component $\sum\nolimits_{m \in S_{jl1}} \beta_{jlm} \bX_m$ of the mean vector $\bX_{A_l^c} \bbeta_{j,l}$ in the univariate linear regression model (\ref{009}) as part of the error vector in the technical analysis for the scaled Lasso. A key observation is that all the error bounds and probability bounds used in the arguments for proving Lemma \ref{Lem1} hold uniformly over the submodels $\mathcal{M}_{jl,\zeta}$. Thus an application of the proof of Lemma \ref{Lem1} shows that with probability $1 - o\{p^{-(\delta - 2)}\}$ tending to one, it holds uniformly over all nodes $j$ in the index sets $A_l$ with $1 \leq l \leq L$ and all submodels $\mathcal{M}_{jl,\zeta}$ that
\begin{align}
\label{084}
& \left\|\hbbeta^*_{j,l, \mathcal{M}_{jl,\zeta}} - \bbeta_{j,l, \mathcal{M}_{jl,\zeta}}\right\|_1 = O(K \lambda), \\
\label{085}
& n^{-1} \left\|\bX_{\mathcal{M}_{jl,\zeta}} (\hbbeta^*_{j,l, \mathcal{M}_{jl,\zeta}} - \bbeta_{j,l, \mathcal{M}_{jl,\zeta}})\right\|_2^2 = O(K \lambda^2),
\end{align}
where $\hbbeta^*_{j,l}$ denotes the SIS-SLasso estimator, which is the scaled Lasso estimator $\hbbeta_{j,l}$ as defined in (\ref{010}) with zero components for $\bbeta$ outside the reduced index set $\mathcal{M}_{jl,\zeta}$ obtained by the SIS, and $\mathcal{M}_{jl,\zeta}$ in the subscripts indicates the corresponding subvectors or submatrices.

In view of (\ref{082}), the intersection of the event $\mathcal{H}_1 \cap \mathcal{H}_2$ and the one given in (\ref{084})--(\ref{085}) still has large probability $1 - o\{p^{-(\delta - 2)}\}$. On such an event, it follows immediately from the sure screening property of $S_{jl0} \subset \mathcal{M}_{jl,\zeta}$, (\ref{084}), and (\ref{083}) that
\begin{equation} \label{086}
\left\|\hbbeta^*_{j,l} - \bbeta_{j,l}\right\|_1 = O(K \lambda).
\end{equation}
Note that the proof of Theorem 2 in  \cite{Lv13}
applies equally well for the largest singular value to show that
\begin{equation} \label{087}
P\left\{\max_{|\Lambda| \leq \widetilde{K}} \lambda_{\max}(n^{-1} \bX_\Lambda\t \bX_\Lambda) \leq O(1)\right\} \leq p^{\widetilde{K}} e^{-C n},
\end{equation}
where $\widetilde{K}$ is as defined in the proof of Lemma \ref{Lem1} and $C$ is some positive constant. Since $\widetilde{K} \leq \widetilde{c}_0 n/(\log p)$ for some sufficiently small positive constant $\widetilde{c}_0$, it is easy to derive that (\ref{087}) entails
\begin{equation} \label{088}
P\left\{\max_{|\Lambda| \leq \widetilde{K}} \lambda_{\max}(n^{-1} \bX_\Lambda\t \bX_\Lambda) \leq O(1)\right\} = 1 - o\left\{p^{-(\delta - 2)}\right\}.
\end{equation}
Thus conditioning on this additional event does not change our asymptotic probability bound $1 - o\{p^{-(\delta - 2)}\}$.

Denote by $\Lambda_0 = \supp(\bbeta_{j,l}) \setminus \mathcal{M}_{jl,\zeta}$. Since  $\|\bbeta_{j,l}\|_0 \leq \widetilde{K}$ as shown in the proof of Lemma \ref{Lem1} which implies $|\Lambda_0| \leq \widetilde{K}$, by (\ref{088}), (\ref{083}) in Condition \ref{Cond4}, and $S_{jl0} \subset \mathcal{M}_{jl,\zeta}$ we have
\begin{align} \label{089}
n^{-1}\|\bX_{\Lambda_0} \bbeta_{j,l,\Lambda_0}\|_2^2 & \leq \lambda_{\max}(n^{-1} \bX_{\Lambda_0}\t \bX_{\Lambda_0}) \|\bbeta_{j,l,\Lambda_0}\|_2^2 \\
\nonumber
& \leq \lambda_{\max}(n^{-1} \bX_{\Lambda_0}\t \bX_{\Lambda_0}) \|\bbeta_{j,l,S_{jl1}}\|_2^2 \\
\nonumber
& \leq O(1) \cdot O(K \lambda^2) = O(K \lambda^2).
\end{align}
Combining (\ref{085}) and (\ref{089}) leads to
\begin{equation} \label{090}
n^{-1} \left\|\bX_{A_l^c} (\hbbeta^*_{j,l} - \bbeta_{j,l})\right\|_2^2 = O(K \lambda^2).
\end{equation}
In light of (\ref{086}) and (\ref{090}), we have shown that with probability $1 - o\{p^{-(\delta - 2)}\}$ tending to one, it holds uniformly over all nodes $j$ in the index sets $A_l$ with $1 \leq l \leq L$ that the same bounds as (\ref{017})--(\ref{018}) in Lemma \ref{Lem1} are also valid for the SIS-SLasso estimator. Therefore, the same arguments as in the proof of Theorem \ref{Thm1} carry through.

\subsection{Proof of Theorem \ref{Thm4}} \label{A.4}
Theorem \ref{Thm4} holds immediately as a consequence of Theorems \ref{Thm2} and \ref{Thm3}.

\section{Proofs of technical results} \label{SecB}

\subsection{Lemma \ref{Lem1} and its proof} \label{B.1}

\begin{lemma} \label{Lem1}
Under Condition \ref{Cond1}, with probability $1 - o\{p^{-(\delta - 2)}\}$ tending to one it holds uniformly over all nodes $j$ in the index sets $A_l$ with $1 \leq l \leq L$ and simultaneously that
\begin{align}
\label{017}
& \|\hbbeta_{j,l} - \bbeta_{j,l}\|_1 = O(K \lambda), \\
\label{018}
& n^{-1} \|\bX_{A_l^c} (\hbbeta_{j,l} - \bbeta_{j,l})\|_2^2 = O(K \lambda^2), \\
\label{019}
& \|n^{-1} \bX_{A_l^c}\t \bE_{j,l}\|_\infty = O(\lambda),
\end{align}
where $\htheta_{j,l} = n^{-1} \widehat{\bE}_{j,l}\t \widehat{\bE}_{j,l}$, $\widetilde{\theta}_{j,l} = n^{-1} \bE_{j,l}\t \bE_{j,l}$, and the additional subscript $l$ indicates the same scalars and vectors as defined previously with the index set $A$ replaced by $A_l$.
\end{lemma}

\textit{Proof of Lemma \ref{Lem1}}. Let us first make a few observations. First, for each index set $A_l$, the random error vector $\bet_{A_l}$ in the scalar form of the multivariate linear regression model (\ref{004}) with index set $A = A_l$ is Gaussian with mean $\bzero$ and covariance matrix $\Ome_{A_l}^{-1}$ and independent of $\bx_{A_l^c}$. Since by Condition \ref{Cond1}, the spectrum of the precision matrix $\Ome$ is bounded between $M^{-1}$ and $M$. We see immediately that the spectrum of its principal submatrix $\Ome_{A_l}$ is also bounded between $M^{-1}$ and $M$, so is that of its inverse $\Ome_{A_l}^{-1}$. This shows that for each corresponding univariate linear regression model (\ref{009}), its error vector $\bE_{j,l}$ is $N(\bzero, \theta_{j,l} I_n)$ with marginal variance $\theta_{j,l}$ bounded between $M^{-1}$ and $M$, where the additional subscript $l$ indicates the same scalars and vectors as defined previously with the index set $A$ replaced by $A_l$.

Second, by Condition \ref{Cond1}, the precision matrix $\Ome$ is $K$-sparse, that is, each of its row or column has at most $K$ nonzero off-diagonal entries. Since $\max_{l} |A_l| = O(1)$, it follows that the total number of nonzero entries $\widetilde{K}$ in the submatrix $\Ome_{A_l^c, A_l}$ is bounded from above by $K |A_l| = O(K)$. In view of $K \leq c_0 n/(\log p)$ for some sufficiently small positive constant $c_0$, we have $\widetilde{K} \leq \widetilde{c}_0 n/(\log p)$ with $\widetilde{c}_0 = O(c_0)$ still some sufficiently small positive constant. Thus for each index set $A_l$, the regression coefficient matrix $\bC_{A_l} = -\Ome_{A_l^c, A_l} \Ome_{A_l}^{-1}$ in the matrix form of the multivariate linear regression model (\ref{005}) with index set $A = A_l$ satisfies that each column vector has at most $\widetilde{K}$ nonzero components. This shows that for each corresponding univariate linear regression model (\ref{009}), its regression coefficient vector $\bbeta_{j,l}$ has sparsity $\|\bbeta_{j,l}\|_0 \leq \widetilde{K} = O(K) \leq \widetilde{c}_0 n/(\log p)$ uniformly over all nodes $j$ and index sets $A_l$.

Third, for each index set $A_l$, the corresponding univariate linear regression model (\ref{009}) is a linear regression model with Gaussian design matrix $\bX_{A_l^c}$ and Gaussian error vector $\bE_{j,l}$ that is independent of $\bX_{A_l^c}$. Note that in light of $\bX = (\bx_1, \cdots, \bx_n)\t$ and (\ref{001}), $\bX_{A_l^c} \sim N(\bzero, I_n \otimes \Sig_{A_l^c})$, where $\Sig_{A_l^c}$ denotes the principal submatrix of $\Sig$ given by the index set $A_l^c$. Since $\Ome$ has spectrum bounded between $M^{-1}$ and $M$, the spectrum of $\Sig = \Ome^{-1}$ is also bounded between $M^{-1}$ and $M$ and so is that of its principal submatrix $\Sig_{A_l^c}$.

Denote by $\mathcal{E}_{j,l}$ the event that the bounds (\ref{017})--(\ref{019}) hold simultaneously for node $j$ in the index set $A_l$. With the above three observations, an application of the proof of Lemma 2 in \cite{RSZZ13}
shows that
\begin{equation} \label{020}
P(\mathcal{E}_{j,l}) = 1 - o\left\{p^{-(\delta - 1)}\right\}.
\end{equation}
Thus applying Bonferroni's inequality over all nodes $j$ in the index sets $A_l$ along with (\ref{020}) yields the uniform bounds (\ref{017})--(\ref{019}) satisfied with probability
\begin{equation} \label{021}
P(\mathcal{E}_1) \geq 1 - p \cdot o\left\{p^{-(\delta - 1)}\right\} = 1 - o\left\{p^{-(\delta - 2)}\right\}
\end{equation}
which converges to one since $\delta \geq 2$, where the event $\mathcal{E}_1$ is defined as
\begin{equation} \label{022}
\mathcal{E}_1 = \bigcap\nolimits_{j \in A_l, 1 \leq l \leq L}\mathcal{E}_{j,l}.
\end{equation}
In view of $\widehat{\bE}_{j,l} = \bX_j - \bX_{A_l^c} \hbbeta_{j,l}$, the fact that $\htheta_{j,l} = n^{-1} \widehat{\bE}_{j,l}\t \widehat{\bE}_{j,l}$ follows easily from the definition of the minimizer $(\hbbeta_{j,l}, \htheta_{j,l}^{1/2})$ of the scaled Lasso problem (\ref{010}).

\subsection{Lemma \ref{Lem2} and its proof} \label{B.2}

\begin{lemma} \label{Lem2}
Under Condition \ref{Cond1}, with probability $1 - o\{p^{-(\delta - 2)}\}$ tending to one it holds uniformly over $1 \leq l \leq L$ that
\begin{equation} \label{023}
\|\widehat{\Ome}_{A_l} - \Ome_{A_l}\|_\infty = O\left\{\max\left(K \lambda^2, \lambda\right)\right\},
\end{equation}
where $\|\cdot\|_\infty$ denotes the entrywise $L_\infty$-norm of a given matrix.
\end{lemma}

\textit{Proof of Lemma \ref{Lem2}}. Note that by (\ref{011}) and (\ref{005}), we have the following decomposition of the residual matrix
\begin{equation} \label{024}
\widehat{\bE}_{A_l} = \bX_{A_l} - \bX_{A_l^c} \widehat{\bC}_{A_l} = \bE_{A_l} - \bX_{A_l^c} (\widehat{\bC}_{A_l} - \bC_{A_l}),
\end{equation}
where $\widehat{\bC}_{A_l} = (\hbbeta_{j,l})_{j \in A_l}$ is a $(p - |A_l|) \times |A_l|$ matrix of estimated regression coefficients. Combining (\ref{012}) and (\ref{024}) yields
\begin{equation} \label{025}
\widehat{\Ome}_{A_l}^{-1} - \Ome_{A_l}^{-1} = n^{-1} \widehat{\bE}_{A_l}\t \widehat{\bE}_{A_l} - \Ome_{A_l}^{-1} = \bxi_1 + \bxi_2 + \bxi_3,
\end{equation}
where $\bxi_1 = n^{-1} \bE_{A_l}\t \bE_{A_l} - \Ome_{A_l}^{-1}$, $\bxi_2 = -2 n^{-1} \bE_{A_l}\t \bX_{A_l^c} (\widehat{\bC}_{A_l} - \bC_{A_l})$, and $\bxi_3 = n^{-1} (\widehat{\bC}_{A_l} - \bC_{A_l})\t \bX_{A_l^c}\t \bX_{A_l^c} (\widehat{\bC}_{A_l} - \bC_{A_l})$. Let us first consider the last two terms $\bxi_2$ and $\bxi_3$ conditional on the event $\mathcal{E}_1$ defined in (\ref{022}). On the event $\mathcal{E}_1$, bounds (\ref{019}) and (\ref{017}) control the maximum rowwise $L_\infty$-norm of matrix $n^{-1} \bE_{A_l}\t \bX_{A_l^c}$ and maximum columnwise $L_1$-norm of matrix $\widehat{\bC}_{A_l} - \bC_{A_l}$, respectively, which lead to
\begin{equation} \label{026}
\|\bxi_2\|_\infty = O(K \lambda^2),
\end{equation}
where $\|\cdot\|_\infty$ denotes the entrywise $L_\infty$-norm of a given matrix. An application of the Cauchy-Schwarz inequality along with bound (\ref{018}) results in
\begin{equation} \label{027}
\|\bxi_3\|_\infty = O(K \lambda^2).
\end{equation}
Note that bounds (\ref{026}) and (\ref{027}) are uniform over $1 \leq l \leq L$. It remains to consider the first term $\bxi_1$.

As mentioned in the proof of Lemma \ref{Lem1}, the spectrum of $\Ome_{A_l}^{-1}$ is bounded between $M^{-1}$ and $M$. In view of (\ref{005}) and (\ref{004}), $n^{-1} \bE_{A_l}\t \bE_{A_l}$ is the oracle sample covariance matrix estimator for $\Ome_{A_l}^{-1}$. Thus the concentration bounds in 
\cite{SS91} and \cite{BL08a}, together with Bonferroni's inequality and $\max_{l} |A_l| = O(1)$, yield for any $t \leq \alpha$,
\begin{equation} \label{028}
P\left\{\|\bxi_1\|_\infty \leq t\right\} = 1 - O(e^{-c n t^2}),
\end{equation}
where $c$ and $\alpha$ are some positive constants. Taking $t = [\delta (\log p)/(cn)]^{1/2}$ in (\ref{028}) and applying Bonferroni's inequality over $1 \leq l \leq L$ lead to
\begin{equation} \label{029}
P(\mathcal{E}_2) \geq 1 - p \cdot O(e^{-c n t^2}) = 1 - O\left\{p^{-(\delta - 1)}\right\} = 1 - o\left\{p^{-(\delta - 2)}\right\},
\end{equation}
where the event $\mathcal{E}_2$ is defined as
\begin{equation} \label{030}
\mathcal{E}_2 = \left\{\max_{1 \leq l \leq L} \left\|n^{-1} \bE_{A_l}\t \bE_{A_l} - \Ome_{A_l}^{-1}\right\|_\infty \leq t = O(\lambda)\right\}.
\end{equation}
Therefore, combining (\ref{025})--(\ref{028}) and (\ref{029}) leads to
\begin{equation} \label{033}
P\left\{\max_{1 \leq l \leq L}\left\|\widehat{\Ome}_{A_l}^{-1} - \Ome_{A_l}^{-1}\right\|_\infty = O\left\{\max\left(K \lambda^2, \lambda\right)\right\}\right\} = 1 - o\left\{p^{-(\delta - 2)}\right\}.
\end{equation}
We still need to derive the bounds for the matrices $\widehat{\Ome}_{A_l}$.

Let us work with the bound $\|\widehat{\Ome}_{A_l}^{-1} - \Ome_{A_l}^{-1}\|_\infty = O\{\max(K \lambda^2, \lambda)\}$. Since $|A_l| = O(1)$, the Frobenius norm $\|\widehat{\Ome}_{A_l}^{-1} - \Ome_{A_l}^{-1}\|_F = O\{\max(K \lambda^2, \lambda)\}$. In light of Condition \ref{Cond1}, the quantity $O\{\max(K \lambda^2, \lambda)$ is bounded above by some sufficiently small positive constant. Then it follows from the matrix perturbation theory (Corollary 6.3.8 of \cite{HJ90}) that
\begin{align*}
\lambda_{\min}(\widehat{\Ome}_{A_l}^{-1}) & \geq \lambda_{\min}(\Ome_{A_l}^{-1}) - \|\widehat{\Ome}_{A_l}^{-1} - \Ome_{A_l}^{-1}\|_F \\
& \geq M^{-1} - O\left\{\max\left(K \lambda^2, \lambda\right)\right\} \geq (2 M)^{-1}
\end{align*}
for large enough $n$. The above spectral inequality leads to $\lambda_{\max}(\widehat{\Ome}_{A_l}) = \lambda_{\min}^{-1}(\widehat{\Ome}_{A_l}^{-1}) = O(1)$. Similarly, we can show that $\lambda_{\min}(\widehat{\Ome}_{A_l})$ is also bounded away from zero.

Note a fact that the entrywise $L_\infty$-norm of any symmetric positive definite matrix is bounded above by its largest eigenvalue. This claim follows from the facts that each diagonal entry is positive and no larger than the largest eigenvalue and that the $2 \times 2$ principal submatrix corresponding to each off-diagonal entry is necessarily nonsingular. Since both $\Ome_{A_l}$ and $\widehat{\Ome}_{A_l}$ have spectra bounded away from $0$ and $\infty$, we see that $\|\Ome_{A_l}\|_\infty = O(1)$ and $\|\widehat{\Ome}_{A_l}\|_\infty = O(1)$, which along with $\max_{1 \leq l \leq L}\|\widehat{\Ome}_{A_l}^{-1} - \Ome_{A_l}^{-1}\|_\infty = O\{\max(K \lambda^2, \lambda)\}$ and $\max_{l} |A_l| = O(1)$ entails
\begin{align}
\label{034}
\left\|\widehat{\Ome}_{A_l} - \Ome_{A_l}\right\|_\infty & = \left\|\Ome_{A_l} \left(\widehat{\Ome}_{A_l}^{-1} - \Ome_{A_l}^{-1}\right) \widehat{\Ome}_{A_l}\right\|_\infty = O\left\{\max(K \lambda^2, \lambda)\right\}.
\end{align}
Therefore, combining (\ref{021}), (\ref{029}), and (\ref{033})--(\ref{034}) yields
\begin{equation} \label{035}
P(\mathcal{E}) = 1 - o\left\{p^{-(\delta - 2)}\right\},
\end{equation}
where the event $\mathcal{E}$ is defined as
\begin{equation} \label{036}
\mathcal{E} = \mathcal{E}_1 \cap \mathcal{E}_2 \cap \left\{\max_{1 \leq l \leq L} \left\|\widehat{\Ome}_{A_l} - \Ome_{A_l}\right\|_\infty = O\left\{\max(K \lambda^2, \lambda)\right\}\right\}.
\end{equation}
Hereafter we condition on the event $\mathcal{E}$.

\subsection{Proof of Proposition \ref{Prop1}} \label{B.3}
For any $\Ome \in \mathcal{G}(M, K)$, we know that each row of $\Ome$ has at most $K + 1$ nonzero components and the spectrum of $\Ome$ is bounded between $M^{-1}$ and $M$. Thus it follows easily that for $\Sig = \Ome^{-1}$ and any $\bu \neq \bzero$,
\begin{equation} \label{048}
\|\bu\|_\infty = \|\Ome \Sig \bu\|_\infty \leq \|\Ome\|_{\infty, \infty} \|\Sig \bu\|_\infty,
\end{equation}
where $\|\cdot\|_{\infty, \infty}$ denotes the operator norm of a matrix induced by the $L_\infty$-norm. Note that $\|\Ome\|_{\infty, \infty}$ is the maximum rowwise $L_1$-norm of $\Ome$, which is bounded above by $(K + 1)^{1/2}$ multiplied by the maximum rowwise $L_2$-norm of $\Ome$, thanks to the Cauchy-Schwarz inequality and the fact that each row of $\Ome$ has $L_0$-norm bounded above by $K + 1$. By the definition of the spectral norm, the maximum rowwise $L_2$-norm of $\Ome$ is further bounded above by $\lambda_{\max}(\Ome) \leq M$, which entails
\begin{equation} \label{049}
\|\Ome\|_{\infty, \infty} \leq (K + 1)^{1/2} M.
\end{equation}
Combining (\ref{048})--(\ref{049}) yields the desired bound $\inf\{\|\Sig \bu\|_\infty/\|\bu\|_\infty: \bu \neq \bzero\} \geq (K + 1)^{-1/2} M^{-1}$.

\subsection{Lemma \ref{Lem3} and its proof} \label{B.4}

\begin{lemma} \label{Lem3}
Assume that Conditions \ref{Cond1}--\ref{Cond2} hold and $K^{1 + \alpha} \lambda = o(1)$. Then with probability $1 - o\{p^{-(\delta - 2)}\}$ tending to one it holds uniformly over all nodes $j$ in the index sets $A_l$ with $1 \leq l \leq L$ that the $L_\infty$-norm cone invertibility factor
\begin{equation} \label{050}
F_{\infty, j, l} = \inf\left\{\frac{\|\widehat{\bR}_{j,l} \bu\|_\infty}{\|\bu\|_\infty}: \|\bu_{S_{j,l}^c}\|_1 \leq \xi \|\bu_{S_{j,l}}\|_1 \neq 0
\right\}
\end{equation}
satisfies $F_{\infty, j, l} \geq c_1 F_\infty$, where $c_1 < 1$ is some positive constant, $S_{j,l}$ denotes the support $\supp(\bbeta_{j,l})$, and $\widehat{\bR}_{j,l} = n^{-1} \bY_{A_l^c}\t \bY_{A_l^c}$ with $\bY_{A_l^c}$ the design matrix $\bX_{A_l^c}$ rescaled columnwise to have $L_2$-norm $n^{1/2}$ for each column.
\end{lemma}

\textit{Proof of Lemma \ref{Lem3}}. Let $\bR$ be the correlation matrix corresponding to the covariance matrix $\Sig = (\sigma_{jk})$. Since the spectrum of $\Sig$ is bounded between $M^{-1}$ and $M$ thanks to the same property of $\Ome$, all diagonal entries $\sigma_{jj}$ of $\Sig$ are also bounded between $M^{-1}$ and $M$ and so are all their reciprocals $\sigma_{jj}^{-1}$. Thus the $L_1$-norms and $L_\infty$-norms induced by both linear transformations corresponding to matrices $\bS = \diag\{\sigma_{11}^{1/2}, \cdots, \sigma_{pp}^{1/2}\}$ and $\bS^{-1} = \diag\{\sigma_{11}^{-1/2}, \cdots, \sigma_{pp}^{-1/2}\}$ are equivalent to the original ones. Thus it follows from the identity
\begin{equation} \label{056}
\bR = \bS^{-1} \Sig \bS^{-1}
\end{equation}
that the $L_\infty$-norm cone invertibility factor $F'_\infty$ with $\Sig$ replaced by $\bR$ in (\ref{047}) and the original one $F_\infty$ defined for $\Sig$ are within a constant factor of each other. To simplify the notation, we still write $F'_\infty$ as $F_\infty$ which is implicitly understood as the $L_\infty$-norm cone invertibility factor defined for $\bR$ hereafter.

For each node $j$ in the index set $A_l$, define the population version of the $L_\infty$-norm cone invertibility factor in (\ref{050}) as
\begin{equation} \label{051}
\widetilde{F}_{\infty, j, l} = \inf\left\{\frac{\|\bR_{A_l^c} \bu\|_\infty}{\|\bu\|_\infty}: \|\bu_{S_{j,l}^c}\|_1 \leq \xi \|\bu_{S_{j,l}}\|_1 \neq 0
\right\},
\end{equation}
where $\bR_{A_l^c}$ denotes the principal submatrix of $\bR$ given by the index set $A_l^c$. As mentioned in the proof of Lemma \ref{Lem1}, $|S_{j,l}| = \|\bbeta_{j, l}\|_0 \leq \widetilde{K} = O(K)$, which together with (\ref{047}) defined for $\bR$ and (\ref{051}) leads to
\begin{equation} \label{052}
\widetilde{F}_{\infty, j, l} \geq F_\infty.
\end{equation}
We will show that the empirical version of the $L_\infty$-norm cone invertibility factor $F_{\infty, j, l}$ in (\ref{050}) concentrates around its population counterpart $\widetilde{F}_{\infty, j, l}$ in (\ref{051}) with overwhelming probability.

Using similar arguments to those for proving (\ref{029}) with $t$ chosen to be $[(\delta + 1) (\log p)/(cn)]^{1/2}$, we can show that
\begin{equation} \label{057}
P(\mathcal{F}) = 1 - o\left\{p^{-(\delta -2)}\right\},
\end{equation}
where $\mathcal{F} = \{\|\hSig - \Sig\|_\infty \leq O(\lambda)\}$ with $\hSig = n^{-1}\bX\t \bX$ and $\|\cdot\|_\infty$ denoting the entrywise $L_\infty$-norm of a given matrix. Note that $\widehat{\bR}_{j,l} = n^{-1} \bY_{A_l^c}\t \bY_{A_l^c}$ is simply the principal submatrix $\widehat{\bR}_{A_l^c}$ of the sample correlation matrix
\begin{equation} \label{058}
\widehat{\bR} = \left(\diag\{\hSig\}\right)^{-1/2} \hSig \left(\diag\{\hSig\}\right)^{-1/2}
\end{equation}
given by the index set $A_l^c$. By some standard calculations, we can show that on the event $\mathcal{F}$, it also holds that $\|\widehat{\bR} - \bR\|_\infty \leq O(\lambda)$. This result together with (\ref{057}) yields
\begin{equation} \label{059}
P(\mathcal{F}_1) \geq 1 - o\left\{p^{-(\delta -2)}\right\},
\end{equation}
where the event $\mathcal{F}_1$ is defined as the intersection of events $\mathcal{F}$ and $\{\|\widehat{\bR} - \bR\|_\infty \leq O(\lambda)\}$.

Finally let us do some algebraic calculations conditional on event $\mathcal{F}_1$. On this event, for each $\bu \in \mathbb{R}^{p - |A_l|}$ satisfying $\|\bu_{S_{j,l}^c}\|_1 \leq \xi \|\bu_{S_{j,l}}\|_1 \neq 0$ we have
\begin{align}
\label{060}
\|\widehat{\bR}_{j,l} \bu\|_\infty & = \|\widehat{\bR}_{A_l^c} \bu\|_\infty \geq \|\bR_{A_l^c} \bu\|_\infty - \left\|\left(\widehat{\bR}_{A_l^c} - \bR_{A_l^c}\right) \bu\right\|_\infty \\
\nonumber
& \geq \widetilde{F}_{\infty, j, l} \|\bu\|_\infty - \|\widehat{\bR} - \bR\|_\infty \|\bu\|_1 \\
\nonumber
& \geq \widetilde{F}_{\infty, j, l} \|\bu\|_\infty - O(\lambda) (1 + \xi) \|\bu_{S_{j,l}}\|_1 \\
\nonumber
& \geq \widetilde{F}_{\infty, j, l} \|\bu\|_\infty - O(\lambda) (1 + \xi) |S_{j,l}| \|\bu_{S_{j,l}}\|_\infty \\
\nonumber
& \geq \left[\widetilde{F}_{\infty, j, l} - O(K \lambda) \right] \|\bu\|_\infty,
\end{align}
since $|S_{j,l}| \leq \widetilde{K} = O(K)$. Therefore, combining (\ref{052}), (\ref{059})--(\ref{060}), and the assumption of $K^{1 + \alpha} \lambda = o(1)$ yields $F_{\infty, j, l} \geq c_1 F_\infty$ for some positive constant $c_1 < 1$, uniformly over all nodes $j$ in the index sets $A_l$ with $1 \leq l \leq L$.

%
%
%
%
%
%

\end{document}